\documentclass[aps,pra,reprint,groupedaddress,longbibliography]{revtex4-1}
\usepackage{graphicx}
\usepackage{dcolumn}
\usepackage{color}
\usepackage[normalem]{ulem} 
\usepackage{amsmath,amsfonts,mathtools}
\begin{document}
\title{Josephson parametric reflection amplifier with integrated directionality}
\author{M.~P.~Westig}
\email[]{m.p.westig@tudelft.nl}
\affiliation{Kavli Institute of NanoScience, Delft University of Technology, 
Lorentzweg 1, 2628 CJ Delft, The Netherlands}
\author{T.~M.~Klapwijk}
\email[]{t.m.klapwijk@tudelft.nl}
\affiliation{Kavli Institute of NanoScience, Delft University of Technology, 
Lorentzweg 1, 2628 CJ Delft, The Netherlands}
\date{\today}
\begin{abstract} 
A directional superconducting parametric amplifier in the GHz 
frequency range is designed and analyzed, 
suitable for low-power read-out of microwave kinetic 
inductance detectors employed in astrophysics and 
when combined with a nonreciprocal 
device at its input also for circuit quantum electrodynamics (cQED). 
It consists of an one wavelength long nondegenerate Josephson parametric 
reflection amplifier circuit. The device has two Josephson 
junction oscillators, connected 
via a tailored impedance to an on-chip passive circuit 
which directs the in- to the output port. The 
amplifier provides a gain of $20$~dB over a bandwidth 
of $220$~MHz on the signal as well as on the idler portion of the 
amplified input and the total photon shot noise referred to the input 
corresponds to maximally $\sim 1.3$ photons per second 
per Hertz of bandwidth. We predict 
a factor of four increase in dynamic range 
compared to conventional Josephson parametric amplifiers.
\end{abstract}
\maketitle
\section{Introduction}
Nonlinearities in superconducting devices, such as the nonlinear 
Josephson inductance \cite{josephson1964}, are building blocks for 
parametric amplification. They can be employed in low noise 
Kerr-type nonlinear oscillators, providing three- or four-wave 
mixing interactions \cite{yurke1988, yurke1989} enabling 
degenerate (phase sensitive) or nondegenerate 
(phase preserving) operation \cite{caves1982}. 
In this letter we design and analyze such a parametric 
amplifier with integrated directionality, facilitating emerging 
low-power read-out schemes for microwave kinetic inductance 
microresonator detectors 
(MKID) \cite{day2003, zmuidzinas2012} employed in astrophysics 
instruments. While our amplifier concept is 
directional although reciprocal, it is only one wavelength long 
which suits and further integrates the MKID read-out backend. 
Recent findings suggest that the vacuum noise generated at the input of our 
parametric amplifier does not decrease the sensitivity of the MKID. 
State-of-the-art read-out schemes of this detector use read-out 
photon numbers of many hundreds of millions in order to overcome the 
cryogenic high electron-mobility amplifier noise and often no 
circulators are used between the detector and the read-out 
amplifier \cite{day2003}. It is expected, however, that 
the MKID sensitivity can be further increased by reducing 
the read-out photon number which can be achieved by parametric 
amplifiers \cite{zmuidzinas2012} such as Josephson parametric 
amplifiers (JPA). Eventually this will help to uncover 
fundamental sources of two-level system noise in superconducting 
microresonators of which no microscopic theory yet exists, important 
for detectors but also for quantum information processors.
In order to be practicable in view 
of the complete detection instrument, maybe 
first in small arrays of about hundred MKID detectors, 
the amplifier bandwidth has to be several tens of a 
MHz large, the dynamic range should enable to 
process first up to on average 100 read-out photons and 
the amplifier should be directional, compact in size and 
easy to fabricate.

A second application of our amplifier 
could be in a circuit quantum electrodynamics (cQED)
\cite{devoret2013, clerk2010, 
bienfait2016a, bienfait2016b, barends2014, hofheinz2009, 
hofheinz2008, mallet2009} measurement scheme. In this case 
our amplifier would have to be supplemented by a nonreciprocal 
device at its input to filter out the vacuum noise 
that would otherwise increase the parasitic photon population of a 
quantum sensitive device connected to the input. Also, few other 
parasitic photons could arise at the input of the parametric 
amplifier due to a finite return loss of the amplifier. 
Research on novel nonreciprocal device 
technologies without lossy and possibly disturbing 
magnetic materials, is presently an active field and is 
expected to provide adequate solutions soon. 
Here, techniques are employed from cavity optomechanics 
\cite{barzanjeh2017, bernier2017, peterson2017, fang2017, 
ruesink2016, shen2016, xu2016, hafezi2012} 
over emulation of circulators with parametric active 
devices \cite{kamal2011} or with Wheatstone 
bridge-based superconducting 
LC resonators \cite {kerckhoff2015, chapman2017}. Also, innovative 
techniques in cQED \cite{koch2010, metelmann2015, ranzani2015, 
kamal2014, estep2014} and different 
Josephson parametric converter circuits 
\cite{sliwa2015, abdo2014, abdo2013} as well as 
other directional Josephson circuits \cite{lecocq2017, macklin2015} 
have been realized. Together with the integrated directionality of our 
amplifier, it is very likely that the parasitic photon population of a 
quantum sensitive device connected to the nonreciprocal 
device-parametric amplifier combination 
is effectively reduced. The nonreciprocal device would have to filter out 
only the vacuum noise and not direct additionally the amplified field to the 
measurement chain which is separately achieved by our amplifier concept.
\section{Amplifier Design and Performance}
\subsection{General Concept}
Our amplifier is operated close to its bifurcation point when the 
dynamics are that of a Duffing oscillator, a small input signal change induces a 
large variation in the system dynamics leading to 
amplification \cite{vijay2007, siddiqi2004, siddiqi2005, manucharyan2007}.

\begin{figure}[tb]
\centering
\includegraphics[width=\columnwidth]{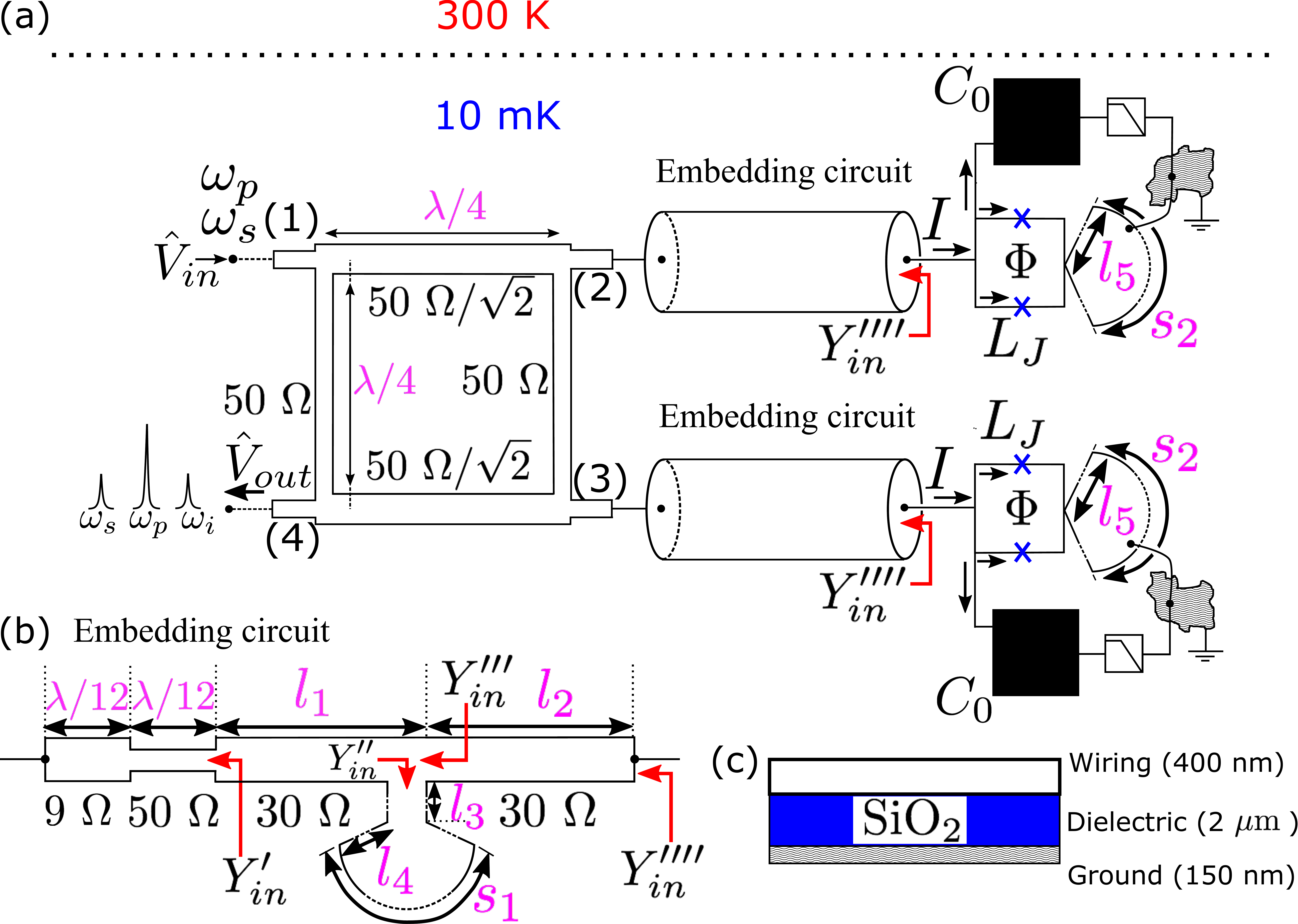}
\caption{\label{fig01}{\bf JPA with directionality}. 
(a) Top view on the 400~nm wiring of the aluminum microstrip circuit, 
consisting of three essential elements; branch-line coupler with 
ports (1) to (4), embedding circuit and Josephson junction oscillators 
with inductance $L_{J}$ (crosses) and capacitance $C_{0}$. A signal 
$\hat{V}_{in}$ having frequency $\omega_{s}$ and 
injected in (1) along with a strong pump tone of frequency $\omega_{p}$ is 
amplified and directed to (4). Additionally, amplified vacuum noise 
is emitted from input port (1) because of the reciprocity of the device. 
Also, few other parasitic photons could 
arise at the input due to a finite return loss of the amplifier. 
(b) Details of the embedding circuit 
of the best performance JPA[-0.36,+0.36], c.f.~Table~\ref{tab01}. 
(c) Cross-section of microstrip layers.}
\end{figure}

Figure~\ref{fig01}(a) shows the aluminum microstrip circuit of 
the JPA. Essential elements are a superconducting 
branch-line coupler (like we have realized experimentally in 
Refs.~\cite{westig2011, westig2012}) 
and an embedding circuit which is connected to two 
individual \emph{rf}-shunted nonlinear 
Josephson junction oscillators (JJO), operated at $\sim 10$~mK. 
The branch-line coupler combines the individual JJOs to a single 
JPA \footnote{We study the functionality of the circuit by simulations in 
\emph{CST} \cite{cst} and find consistence with an analytical model describing our 
microstrip geometry up to an uncertainty of 3\% \cite{supp, kautz1978}, 
assuming a normal-state resistance of 
$\rho_{n} = 0.1~\mu\Omega\mathrm{cm}$ and a superconducting gap of 
$\Delta = 0.18$~meV} and provides signal directivity while the matched input admittance 
$Y_{in}''''$ of the embedding circuit determines gain, noise 
and bandwidth of the JPA via its engineered conductance 
and susceptance portions. The value of $Y_{in}''''$ is dominated 
by a capacitive shunt of the first $\lambda/12$ section 
in the embedding circuit rather than by a higher 
impedance inductive load like realized in earlier work \cite{mutus2014, roy2015}.

Each of the two JJOs should be characterized by the same 
plasma frequency $\omega_{0} = \left((2\pi I_{c}(\Phi))/(\phi_0 C_0)\right)^{-1/2}$. 
Here, $L_{J}(\Phi)=\phi_{0}/(2\pi I_{c}(\Phi))$ is the 
Josephson inductance being $0.12~\mathrm{nH}$ in our case and 
$C_{0} = 4.0~\mathrm{pF}$ is the shunting capacitor. Furthermore, 
$\phi_{0} = h/(2e)$ is the flux quantum and 
$I_{c}(\Phi) = 2i_{c}\lvert\cos\left(\pi \Phi/\phi_{0}\right)\rvert$
is the total critical current for each Josephson SQUID shown in 
Fig.~\ref{fig01}(a) with $i_{c}$ being the individual currents of the 
single junctions in the SQUID. 
An externally applied small magnetic flux bias
induced in the SQUID loop, $\Phi$, tunes 
$\omega_{0}/(2\pi) \sim 7.3~\mathrm{GHz}$ and 
the JJO admittance $Y_{0} = \sqrt{C_{0}/L_{J}(\Phi)}$.
The JJOs are pumped 
through their embedding circuit by a strong coherent tone of frequency 
$\omega_{p}/(2\pi) \sim 6.0~\mathrm{GHz}$ which provides the energy for the amplification 
and by a much weaker quantum signal of frequency $\omega_{s}$ which shall 
be amplified. In this work we consider nondegenerate four-wave mixing as amplifying 
mechanism for which $2\omega_{p} = \omega_{s} + \omega_{i}$; 
two pump photons at $\omega_{p}$ transfer their energy 
into signal and idler modes symmetric around $\omega_{p}$, c.f.~Fig~\ref{fig01}(a).
The Josephson SQUIDs are designed on top of a $2~\mu\mathrm{m}$ 
thick $\mathrm{SiO_{2}}$ dielectric layer, c.f.~Fig.~\ref{fig01}(c). 
Each of the SQUIDs is \emph{rf}-shunted through a virtual short 
around 6~GHz, realized by a broadband radial stub tuner \cite{sorrentino1992} 
with dimensions $\lbrace s_{2}, l_{5} \rbrace$ and characteristic 
input impedance of $50~\Omega$. A bonding wire 
connects the edge of the radial stub directly to an island on the 
ground plane, connecting the otherwise galvanically separated 
microstrip layers. On the 
other side of the SQUID, the same island connects to a planar low 
pass filter (e.g.~a standard microstrip Chebyshev \emph{rf}-blocking 
filter \cite{collin1992}), being connected to 
the oscillator's shunting capacitor $C_{0}$. Therefore, a \emph{dc}-current can flow 
through the SQUID, providing the inductance $L_{J}$, and at the same time 
the bonding wires do not disturb the \emph{rf} circuit.
\begin{table}[t]
\caption{\label{tab01}Three JPA designs resulting in different residual 
frequency dependent imaginary parts in 
$-i\tilde{\omega}_{s} + \kappa[\tilde{\omega}_{s}]/2 \propto Y_{in}''''[\tilde{\omega}_{s}]$ of 
Eq.~(\ref{eq:02}). We quantify the residue by indicating the 
minimum/maximum slope of the imaginary part as index, 
e.g.~JPA[-0.36,+0.36]. A weakly frequency dependent imaginary 
part with a slope symmetric around zero yields a maximal 
bandwidth determined by 
$\mathrm{Re}[Y_{in}''''[\tilde{\omega}_{s}]]$. 
The numbers specify the dimensions of the circuit parts 
in $\mu$m from which the characteristic impedance in $\Omega$ is 
given in curly brackets.}
\begin{ruledtabular}
\begin{tabular}{llll}
\textrm{Elements}&
\textrm{JPA[-1,+1]}&
\textrm{JPA[-0.36,+0.36]}&
\textrm{JPA[-0.53,-0.19]}\\
\colrule
$\lambda/4$ [(1)-(2)]\footnote{The connection of ports (3) and (4) has the same length.} 
& 7098 $\lbrace 50/\sqrt{2} \rbrace$ & 7098 $\lbrace 50/\sqrt{2} \rbrace$ & 7098 $\lbrace 50/\sqrt{2} \rbrace$ \\
$\lambda/4$ [(1)-(4)]\footnote{The connection of ports (2) and (3) has the same length.} 
& 7306 $\lbrace 50 \rbrace$ & 7306 $\lbrace 50 \rbrace$& 7306 $\lbrace 50 \rbrace$ \\
$\lambda/12$ [1st] & 1177 $\lbrace 5 \rbrace$ & 1522 $\lbrace 9 \rbrace$ & 1941 $\lbrace 16 \rbrace$\\
$\lambda/12$ [2nd] & 1332 $\lbrace 50 \rbrace$ & 1722 $\lbrace 50 \rbrace$ & 2077 $\lbrace 50 \rbrace$\\
$l_{1}$ & 1587 $\lbrace 30 \rbrace$ & 2137 $\lbrace 30 \rbrace$ &2650 $\lbrace 30 \rbrace$\\
$l_{2}$ & 8815 $\lbrace 30 \rbrace$ & 9515 $\lbrace 30 \rbrace$ &10262 $\lbrace 30 \rbrace$\\
$l_{3}$ & 262 $\lbrace 30 \rbrace$ & 182 $\lbrace 30 \rbrace$ &182 $\lbrace 30 \rbrace$\\
$l_{4}$ & 290 & 224 & 155 \\
$s_{1}$ & 658  & 553 & 422\\
$l_{5}$ & 2598 & 2598 & 2598\\
$s_{2}$ & 4848 & 4848 & 4848
\end{tabular}
\end{ruledtabular}
\end{table}
\subsection{Equation of Motion, Gain and Noise}
The equation of motion (EOM) for each of the two 
JJOs is described by independent 
Duffing equations obtained from the RCSJ-model \cite{tinkham2004} 
and by \emph{Kirchhoff's} law applied to 
the equivalent circuit of the JJO and its embedding circuit, 
shown in Fig.~\ref{fig01}(a) and (b):
\begin{equation}
\begin{split}
\label{eq:01}
\ddot\delta(t) + \kappa \dot \delta(t) 
+ \omega_{0}^{2}\left[\delta(t) - \frac{\delta(t)^{3}}{6}\right] 
- \omega_{0}^{2} \frac{I_{p}(t)}{I_{c}(\Phi)}
= \frac{4\pi \hat{I}_{in}(t)}{\phi_{0}C_{0}}~.
\end{split}
\end{equation}
This equivalent circuit describes the pumping of the particular 
JJO from a (parallel) current source with admittance $Y_{in}''''$. 
Furthermore, $\delta(t) = \delta_{p}(t) + \hat{\delta}_{s}(t)$ is the 
intra-oscillator field and consists of a classical term $\delta_{p}(t)$ due 
to the coherent large amplitude pump tone, 
$I_{p}(t) = \bar{I}_{p}\cos\left(\omega_{p} t\right)$, and a quantum 
term $\hat{\delta}_{s}(t)$ induced by the weak signal $\hat{I}_{in}(t)$. 
The dissipation is quantified by a rate 
$\mathrm{Re}[\kappa] = \mathrm{Re}[Y_{in}'''']/C_{0}$, obtaining 
the convenient value $\mathrm{Re}[\kappa]/(2\pi) \sim 1.33$~GHz.
Equation~(\ref{eq:01}) only holds for a strong nonlinearity of the 
JJO compared to the linear inductance $L_{env}$ 
contributed by the embedding circuit. This is quantified by the participation ratio 
$p = L_{\parallel}/L_{J}$ \cite{manucharyan2007} and $L_{\parallel}^{-1} = L_{env}^{-1} + L_{J}^{-1}$
where $L_{\parallel}$ is the total parallel inductance. We find that $L_{env}$ is larger by about a 
factor 22 compared to $L_{J}$ so that the nonlinearity of the JJO is strong enough 
to assume the ideal case provided by Eq.~(\ref{eq:01}), where the nonlinear term reads 
$p\delta(t)^{3}/6 \approx \delta(t)^{3}/6$.
\begin{figure}[tb]
\centering
\includegraphics[width=\columnwidth]{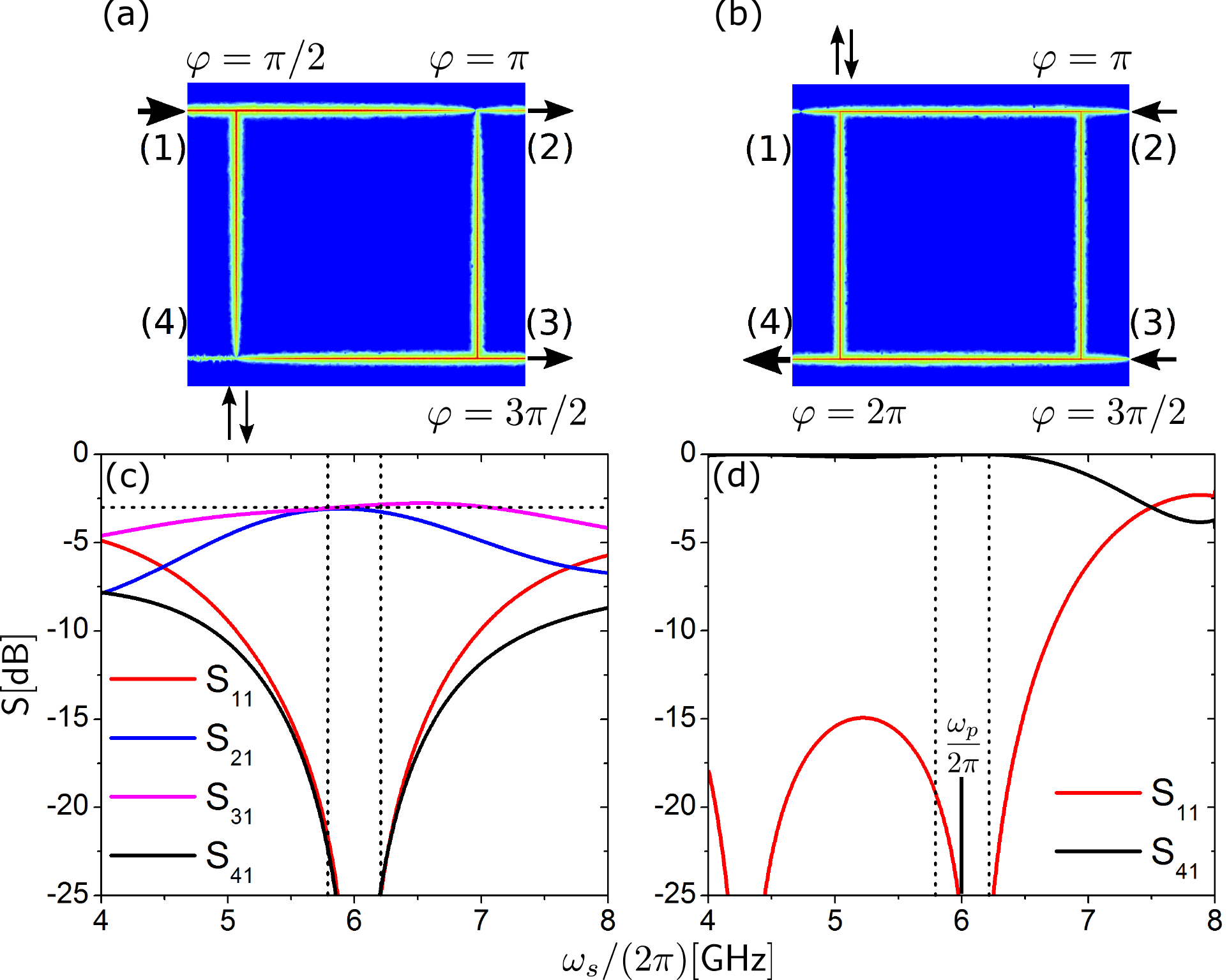}
\caption{\label{fig02}{\bf Directionality}. (a) and (b) show a 
snapshot of $\lvert {\bf E}({\bf r}) \rvert$ on the branch-line 
coupler at the design frequency 6~GHz. (c) and (d) quantify the 
corresponding scattering parameters, determining 
a slight asymmetry in the coupler performance. Note that 
in (d) the scattering parameters are fully reciprocal, in 
particular $S_{41} = S_{14}$. Hence, vacuum noise is generated 
at input port (1). In applications for cQED this noise has to be 
filtered out by an additional nonreciprocal device connected to input port (1).
(a) Excitation of port (1) and phase delayed distribution 
to ports (2) and (3) connecting to the amplifier circuits (not shown); 
port 4 receives no signal and this is indicated by the up-down arrows illustrating 
an out-of-phase condition. (b) Returning phase delayed signals 
from the amplifier circuits, combining this time in-phase at the 
output port (4) and not at port (1), quantified by the value $S_{11}$. 
In (b) the gain is set to $G = 1$ for illustrative purposes. 
Vertical dashed lines are taken from Fig.~\ref{fig04}(b) indicative of 
the amplifier $-3$~dB bandwidth and the horizontal dashed line is the $-3$~dB level 
of the branch-line coupler.}
\end{figure}
For $\hat{I}_{in} = 0$, 
one obtains the steady state solution of Eq.~(\ref{eq:01}) which determines the classical 
pump intra-oscillator field $\delta_{p}(t)$ with maximum amplitude $\delta_{p,max}$. 

The case $\hat{I}_{in} \not= 0$ obtains the quantum 
intra-oscillator field by subtracting the steady state solution from Eq.~(\ref{eq:01}). 
We proceed by making a Fourier transformation and subsequent transition into the 
rotating frame \cite{supp} of the pumping field at frequency 
$\omega_{p}$, obtaining:
\begin{equation}
\label{eq:02}
\left[i(\tilde{\Omega}_{p} -
\tilde{\omega}_{s}) + \frac{\kappa[\tilde{\omega}_{s}]}{2}\right]
\hat{a}_{s}[\tilde{\omega}_{s}]
- \frac{i\omega_{0}\delta_{p,max}^{2}}{16}
\hat{a}_{i}^{\dagger}[-\tilde{\omega}_{s}]
= \hat{a}_{in}[\tilde{\omega}_{s}]~.
\end{equation}
We denote with $\tilde{\Omega}_{p} = \omega_{0} - \omega_{p} - \omega_{0}\delta_{p,max}^{2}/8$ 
the effective pump frequency detuning and with $\tilde{\omega}_{s} = \omega_{p} - \omega_{s}$ 
the signal frequency detuning. 
The intra-oscillator field operators $\hat{a}_{s,i}$ describing the signal and 
idler modes are related up to a phase factor to $\hat{\delta}_{s}$. 
They obey the standard Heisenberg 
EOM of the nondegenerate JPA \cite{gardiner1985, laflamme2011} 
through the four-wave mixing Hamiltonian $\hat{H} = 
\hbar \tilde{\Omega}_{p}(\hat{a}_{s}^{\dagger}\hat{a}_{s} + \hat{a}_{i}^{\dagger}\hat{a}_{i}) 
+i\hbar (g/2) (\hat{a}_{s}^{\dagger}\hat{a}_{i}^{\dagger} - \hat{a}_{s}\hat{a}_{i})$ 
where $g=2\Lambda \bar{n}$ is a Kerr-like nonlinearity \cite{laflamme2011, clerk2010} 
with $\Lambda \bar{n} \sim \omega_{0}\delta^{2}_{p,max}/16$ being the 
product of oscillator nonlinearity and the large average number of quanta 
$\bar{n}\gg 1$ in the oscillator. 
\begin{figure}[t]
\centering
\includegraphics[width=\columnwidth]{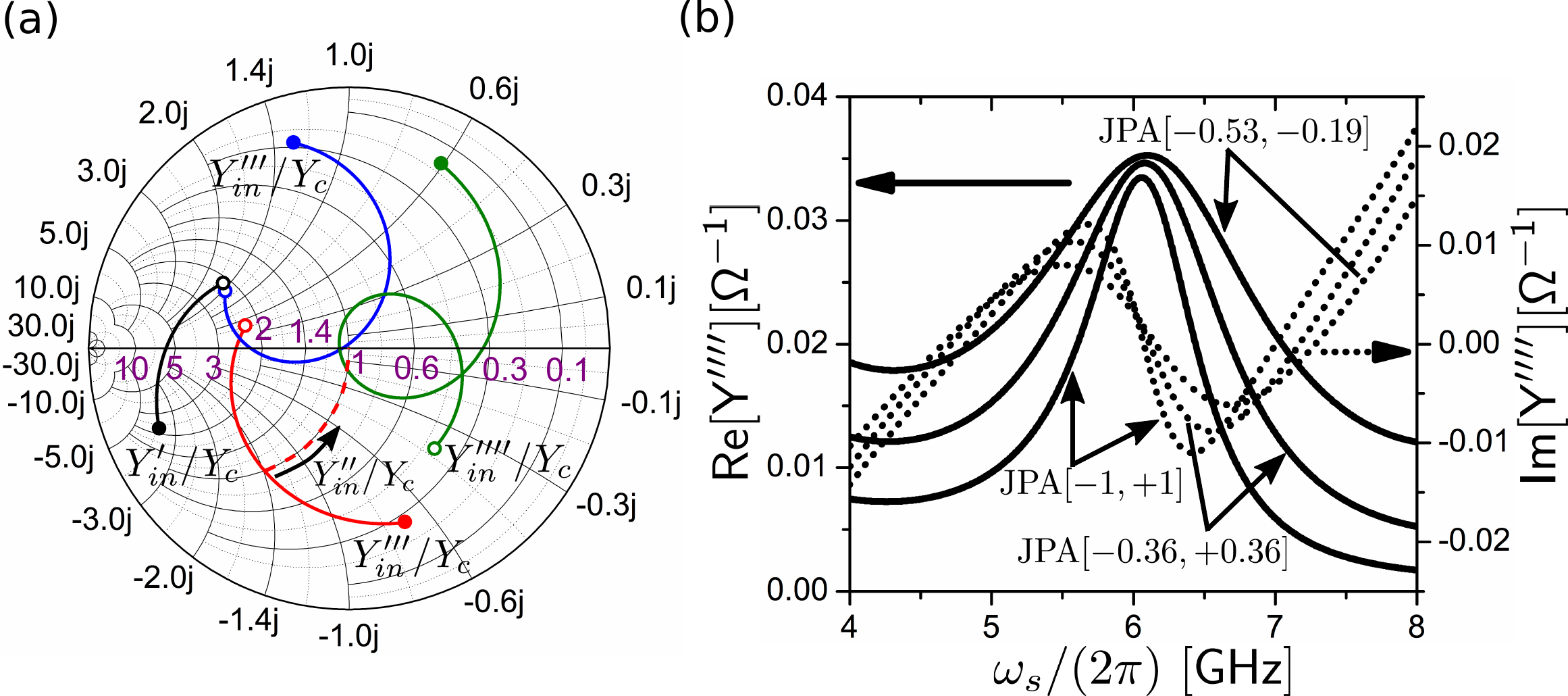}
\caption{\label{fig03}{\bf Embedding circuit design}. 
(a) The Smith chart shows the design relevant input admittances in the polar plane, 
normalized to the characteristic admittance of the particular circuit part, 
extracted from Table~\ref{tab01} for the case of JPA[-0.36,+0.36]. 
The shunting capacitance which is encoded in the trace 
$Y_{in}'$ determines the shape of the desired admittance $Y_{in}''''$, controlling the 
amplifier performance. The red and blue traces are the admittances before and 
after shunting with the stub with admittance $Y_{in}''$, pulling the red trace into 
the vicinity of the desired green trace.
The open/closed circles indicate the start(4~GHz)/stop(8~GHz) 
frequency. (b) Separate real (conductance) and imaginary (susceptance) 
components of $Y''''_{in}[\omega_{s}/(2\pi)]$ for a single embedding circuit used in 
the three designs.}
\end{figure}
The solution of the Heisenberg EOM gives the oscillator 
susceptibility, linking the intra-oscillator field to the input field. 
The inverse susceptibility matrix $\chi^{-1}$ is the coefficient matrix of 
Eq.~(\ref{eq:02}) and its adjoint equation; 
$\chi[\tilde{\omega}_{s}]^{-1}\cdot\hat{\bf{a}}[\tilde{\omega}_{s}] = 
\hat{\bf{a}}_{in}[\tilde{\omega}_{s}]$ where 
$\hat{\mathbf{a}}_{in} = \left(\hat{a}_{in}[\tilde{\omega}_{s}], 
\hat{a}_{in}^{\dagger}[-\tilde{\omega}_{s}]\right)^{T}$ 
and
$\hat{\mathbf{a}} = \left(\hat{a}_{s}[\tilde{\omega}_{s}], 
\hat{a}_{i}^{\dagger}[-\tilde{\omega}_{s}]\right)^{T}$ \cite{clerk2010, roy2015}.
By inverting $\chi^{-1}$ and evaluating the element $\chi_{11}$, 
we determine the photon number gain 
$G_{s}[\tilde{\omega}_s] = \mathcal{C}_{1} \lvert1 - 
\mathrm{Re}[\kappa[\tilde{\omega}_{s}]]\chi_{11}[\tilde{\omega}_{s}]\rvert^{2}$ which can be 
understood as a reflection coefficient at the JJO larger than one and $\mathcal{C}_{1}$ is a 
circuit dependent correction factor. A similar equation holds for the idler field. 
Knowing the gain, we can estimate the noise added by the nondegenerate 
JPA by using Ref.~\cite{laflamme2011}. We modify their result to account for the complete noise 
referred to the input of the amplifier, consisting of minimum
half a photon of shot noise per second and per Hz of bandwidth 
amplified from the signal and 
also from the idler field; in total $T_{N}[\tilde{\omega}_{s}] = \mathcal{C}_{2}^{-1}
(\hbar \omega_{p}/k_{B} - (2\hbar\tilde{\omega}_{s}/k_{B})
\{(1/4 + (G[0]/3)(\omega/(1+\omega^2))^2)^{1/2} - 
(G[0]/3)^{1/2}\omega/(1+\omega^{2})]\}$. Here, $\mathcal{C}_{2}$ is another 
circuit dependent correction factor and 
$\omega = \tilde{\omega}_{s} \lvert\chi_{11}[0]\rvert$. For $\tilde{\omega}_{s} = 0$ 
this is the fundamental result of the Haus-Caves theorem \cite{haus1962, caves1982} and
for $\tilde{\omega}_{s} \not= 0$ we obtain 
an approximate relation of the noise in our amplifier for small detuning.
\subsection{Embedding circuit}
We now describe how the two independent JJOs function 
as one single JPA via their surrounding circuit, c.f.~Fig.~\ref{fig01}(a). The strong 
coherent pump tone is applied together with the weak quantum signal 
to the same input port (1) of a superconducting 
branch-line coupler. Our amplifier concept contributes a passive and 
ultra low-loss signal routing functionality to the circuit 
toolbox that can in cQED applications 
be combined with existing nonreciprocal circuits 
to filter out parasitic photons that are generated at the input 
port (1) of our amplifier. 
The voltage amplitudes which propagate to 
ports (2), (3) and (4) are described by a scattering relation, 
$\left[\bf{S}\right] \left[\bf{V}^{\rightarrow}\right] = \left[\bf{V}^{\leftarrow}\right]$,
where $\left[\bf{V}^{\rightarrow,\leftarrow}\right]$ 
are the ingoing ($\rightarrow$) and outgoing ($\leftarrow$) 4-component 
voltage waves applied to ports (1)-(4) and $\left[\bf{S}\right] = \frac{1}{\sqrt{2}} 
\begin{psmallmatrix}0&1&i&0\\1&0&0&i\\i&0&0&1\\0&i&1&0\end{psmallmatrix}$ 
is unitary for a lossless ideal branch-line coupler. 
An input voltage $V^{\rightarrow} = 
V_{p}[\omega_{p}]\exp{(-i\theta_{p})} + \hat{V}_{in}[\omega_{s}]\exp{(-i\theta_{s})}$ 
applied only to port (1), 
is divided equally between the ports (2) and (3) together with a \emph{relative} phase 
shift of $\pi/2$. Port (4) receives no signal in this case. 
The divided signals leaving ports (2) and (3) couple via the embedding 
circuits of dimensions $\lbrace l_{i}, s_{i} \rbrace$ to the JJOs 
in which they are amplified and reflected. Ideally, the reflected 
signals still carry this \emph{relative} phase shift of $\pi/2$ and are back-coupled 
to ports (2) and (3). An evaluation of the output voltages 
via the $\left[\bf{S}\right]$-matrix shows that now the divided signals combine again 
constructively at port (4) whereas port (1) receives no signal.
We visualize this effect in Fig.~\ref{fig02} and show that
the routing of the signals in our circuit is entirely passive and not susceptible 
to loss within the amplifier bandwidth. 
For a nondegenerate JPA which preserves the input phase(s) 
at the output \cite{caves1982}, we assume without 
loss of generality $\theta_{p} = \theta_{s} = \pi/2$ 
at port (1). The density plots for the electric field between the ground plane and 
the wiring circuit (c.f.~Fig.~\ref{fig01}(c)), $\lvert {\bf E}({\bf r}) \rvert$, using this choice 
of phases show the directional operation on the branch-line coupler.
The calculated scattering parameter magnitudes for our circuit are 
shown in the same figure. They are related to powers ($\propto V^{2}$) and 
quantify the signal distribution, yielding an almost 
perfect realization of the ideal $\left[\bf{S}\right]$-matrix over the envisioned 
operation bandwidth of the JPA. 
\begin{figure}[tb]
\centering
\includegraphics[width=\columnwidth]{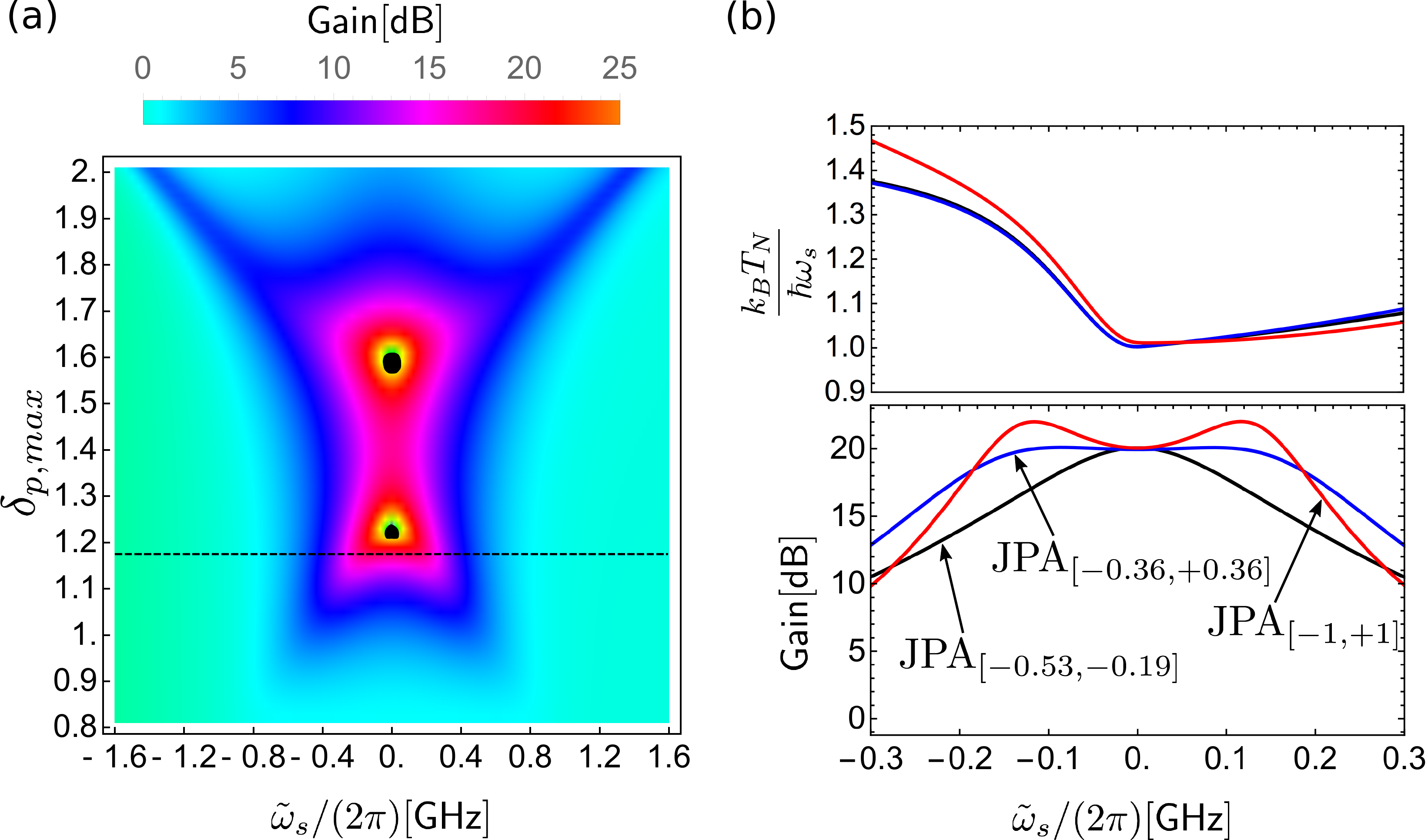}
\caption{\label{fig04}{\bf Amplifier performance}. (a) Power gain as a function of signal frequency 
detuning $\tilde{\omega}_{s}$ and pumping amplitude $\delta_{p,max}$ for 
JPA[-0.36,+0.36]. The black dashed line indicates the optimal (lowest power) pumping 
amplitude to achieve a flat gain profile of about 20~dB. Black regions distinguish 
high-gain regimes $\gg 25$~dB but with significantly decreased bandwidth.  
(b) Approximate shape of the noise temperature referred to the input and 
optimal power gain as a function of 
signal frequency detuning for the three designs.}
\end{figure}

We design the embedding circuit 
by visualizing the voltage reflection 
coefficient and, hence, the normalized complex transmission line input admittance 
$Y_{in}/Y_{c} = \left(Y_{L} + i Y_{c}\tanh\left(\gamma l \right)\right)
/\left(Y_{c} + i Y_{L}\tanh\left(\gamma l \right)\right)$ 
in the polar plane of the Smith chart \cite{smith1939, collin1992}, 
c.f.~Fig~\ref{fig03}(a) and Fig.~\ref{fig01}(a). 
Here, $Y_{c}$ is the characteristic admittance of the particular microstrip 
segment, $Y_{L}$ is the termination admittance, $\gamma$ is the complex 
propagation constant \cite{kautz1978, supp} and $l$ is the length. 
Finally, the complex valued admittance is 
plotted in the Smith chart. A point represents 
the intersection between the corresponding 
conductance circle, where the normalized conductance is indicated 
on the horizontal axis and the susceptance circle, where the normalized 
susceptance is indicated around the Smith chart. For a given admittance, a change 
in $l$ will rotate the trace in the Smith chart by $2\gamma l$ and 
connecting admittances of different values will lead to jumps in the overall admittance 
(compare with the traces in Fig.~\ref{fig03}(a)), 
transforming the admittance and changing the shape of the 
conductance and the slope of the susceptance of $Y_{in}''''$ 
like shown in Fig.~\ref{fig03}(b). The correction factors for 
gain and noise are given by the 
scattering parameters and the coupling to the JJOs,
$\mathcal{C}_{1} = \lvert S_{41}[\tilde{\omega}_{s}]\rvert$ shown in Fig.~\ref{fig02}(d) and 
$\mathcal{C}_{2} = (\lvert S_{21}[\tilde{\omega}_{s}]\rvert 
+ \lvert S_{31}[\tilde{\omega}_{s}]\rvert) \times \mathcal{T}$
where the scattering parameters are the ones shown in Fig.~\ref{fig02}(c) 
and $\mathcal{T}$ is the signal coupling to the JJOs, 
being a factor between '1' and '0.95', c.f.~\cite{supp}. 

The directional signal routing in our device 
relies fundamentally on the imposed relative 
phase difference of $\pi/2$ between the two 
JJOs and we have to estimate the influence 
of slightly detuned JJOs. We find that a relative plasma frequency detuning 
of $200$~MHz will cause an additional phase difference of $\sim \pi/18$ in the 
outgoing amplified signals. While this will only slightly change 
the coupler directivity 
$S_{41}$ in Fig.~\ref{fig02}(d), the return loss 
$S_{11}$ of the device port (1) indicated in the 
same figure will degrade to $-12$~dB from its original value of $<-25$~dB \cite{supp}.
In the aforementioned situation, the two JJOs 
differ also in their gain by about 0.8~dB which has a negligible influence.
\section{Results}
Figure~\ref{fig04} and Table~\ref{tab01} 
summarize our results for gain, noise and the designs of the 
JPA. While we obtain the best performance for JPA[-0.36,+0.36], 
the other two designs show the influence of the real and 
imaginary part of $Y_{in}''''$ on the amplifier performance. 
For design JPA[-0.36,+0.36], the term 
$-\tilde{\omega}_{s} + \mathrm{Im}[\kappa[\tilde{\omega}_{s}]]$ 
in Eq.~(\ref{eq:02}) assumes the smallest frequency dependence 
with a slope symmetric around zero, maximizing the amplification 
bandwidth. For the other two designs the same term contains a 
much stronger frequency dependence and the slope is not 
symmetric around zero, resulting in a decreased performance, 
therefore, bandwidth of the amplifier. The dynamic range for 
a single Josephson junction oscillator operated close to the 
bifurcation point scales with 
$P_{dyn} \propto I_{c}^{2}/Q$ \cite{manucharyan2007}, where 
$Q \approx \omega_{0} \mathrm{Re}\left[Z_{in}''''\right] C_0$. 
The two JJOs in our circuit effectively 
double the critical current which increases the dynamic 
range by a factor of four.
 
It can be further increased by increasing the current density of the 
Josephson junctions which is rather limited for 
Al/AlOx/Al junctions. Higher values of up to $J_{c} = 78~\mathrm{kA/cm^2}$ 
are reachable with AlN barriers in Nb-based circuits \cite{zijlstra2007} which would 
increase the dynamic range by up to three orders of magnitude 
compared to existing JPAs and would enable to read-out 
large arrays of detectors containing some thousands of pixels \cite{day2003} 
or multiple qubits \cite{barends2014}. 
In this amplifier technology, however, the SQUID cannot be fabricated anymore using the well 
established angle-evaporation technique and one has to rely on 
trilayer Josephson junctions \cite{westig2011, westig2012}.
\section{Discussion}
We have designed and analyzed 
a broadband and compact JPA with integrated directionality 
which adds only about one single photon of total noise at the 
input. While our proposed device is fully reciprocal, 
nonreciprocity can be achieved by combining our device 
with existing nonreciprocal devices at the input of our amplifier.
Employing existing high-current density Josephson 
junctions would increase the dynamic range significantly 
compared to existing JPAs. Our embedding circuit is 
general enough to tune the gain 
of signal and idler modes independently, providing 
interesting opportunities to tailor nonclassical microwave 
light \cite{armour2015, westig2017}.

In closing, we address two specific examples, 
where our proposed amplifier adds improved functionality.

In cQED, an increasing number of experiments reads out 
a cavity state by using a one-port JPA together with a 
nonreciprocal device, which directs the amplified field 
to the post-processing electronics and protects
the cavity from noise. Commercially available nonreciprocal
magnetic circulators are mainly characterized by
their isolation, which quantifies to what extent the 
circulator can block radiation emitted towards the 
quantum sensitive cavity. A typical isolation value for 
these commercial circulators amounts to -20 dB. 
This is also true for the novel non-magnetic circulator 
reported in \cite{chapman2017}. As a consequence, the amplified 
vacuum noise of such a one-port JPA, emitted towards the 
nonreciprocal device, results in one parasitic photon per 
second and per Hz of bandwidth, transmitted towards the 
cavity. A desired cavity field, amplified by the JPA, will further 
enhance the number of these parasitic photons. Improvement 
over a broad bandwidth is difficult to achieve. 

Our amplifier concept, provides an attractive solution since
it emits only amplified vacuum noise from input
port (1) and emits separately the amplified field and vacuum
noise from output port (4). Therefore, a nonreciprocal
device with a given isolation connected to input
port (1), reduces also the parasitic photons transmitted through 
the nonreciprocal device towards the cavity.

In another example, in astronomical instrumentation 
\cite{baselmans2017, ferrari2018} microresonator arrays 
of 20.000 pixels have been realized recently. 
The dissipation and the noise of the read-out amplifiers is becoming 
a very important limiting factor. In addition, in order to reduce standing 
waves in the read-out signal it would be beneficial to integrate an 
amplifier on the same chip with the MKID array. The fabrication of our 
amplifier is compatible with the currently used MKID technology. 
Our amplifier concept makes it possible to connect the read-out line of 
the MKID array directly to input port (1). The combination of low noise, 
broad-band and integrability make our proposed amplifier very suitable 
for use with MKID arrays.
\begin{acknowledgements}
We acknowledge funding through the European Research 
Council Advanced Grant No.~339306 (METIQUM). 
We have had fruitful discussions with Karl Jacobs and 
Patrick P{\"u}tz of KOSMA, 1.~Physikalisches Institut, Universit{\"a}t 
zu K{\"o}ln, Germany, related to the micro-fabrication of the circuit 
and the various Josephson junction technologies. 
Furthermore, we appreciated helpful comments 
from Udson Mendes, Universit{\'e} de Sherbrooke, Canada, on the general 
content. We would like to thank the referees for critical 
and essential comments on the input noise created by our parametric 
amplifier and for other general comments which helped to improve the 
manuscript.
\end{acknowledgements}
\onecolumngrid
\vspace{\columnsep}
\newpage
\begin{center}
\textbf{\large Josephson parametric reflection amplifier with integrated directionality: Supplemental material}
\end{center}
{\it M.~P.~Westig (m.p.westig@tudelft.nl) and T.~M.~Klapwijk (t.m.klapwijk@tudelft.nl)} \\ \\
{\it Kavli Institute of NanoScience, Delft University of Technology, 
Lorentzweg 1, 2628 CJ Delft, The Netherlands}
\vspace{\columnsep}
\twocolumngrid
\setcounter{equation}{0}
\setcounter{figure}{0}
\setcounter{table}{0}
\setcounter{page}{1}
\makeatletter
\renewcommand{\theequation}{S\arabic{equation}}
\renewcommand{\thefigure}{S\arabic{figure}}
\newcommand\id{\ensuremath{\mathbbm{1}}}
In this supplemental information we provide additions to several aspects 
of the main text which in our believe are worth-while discussing in more detail. 
We solve the equation of motion for the 
single nonlinear Josephson junction oscillators (JJO), both when a strong pump 
tone and a weak quantum signal is applied to their input. 
This yields the gain and noise for the single JJOs when operated as a
Josephson parametric amplifier (JPA). Additionally, we 
provide a suitable input-output formalism 
for the JPA microwave fields in order to quantify the 
parametric gain and noise of the device when two JJOs 
are combined to form a single directional JPA, 
being the topic of the main text.
Also, we add useful material summarizing a 
design and its characterization 
of a dielectric loaded coplanar waveguide-to-microstrip 
transformer circuit. This circuit provides the possibility to 
connect our microstrip JPA to other 
quantum circuits or to the microwave cabling of the 
experiment, often designed in coplanar 
waveguide technology. Finally, we 
provide an analytical model for the dispersion relation on the 
superconducting microstrip transmission line which 
allows to evaluate the design changes of the JPA circuit 
as a function of the aluminum normal-state 
resistivity (treated in this work) 
or as a function of the resistivity of any other superconducting material 
which is chosen to pattern the circuit. 
\section*{Equation of motion for 
the single JJO acting as nondegenerate JPA}
The tone which is pumping the two JJOs through 
ports (2) and (3), shown in Fig.~\ref{fig01}(a) of the main text, can be 
described as a pumping current source connected in parallel to the input 
admittance $Y_{in}''''[\tilde\omega_s]$ of the JJO's electromagnetic (EM) 
environment (acting as the source admittance of the pumping current source)
and to the respective JJO; the JJO is 
attached in a SQUID configuration and acts effectively as a 
magnetic-flux tunable nonlinear oscillator.

For reasons of simplicity, we will consider in the following treatment 
only a single JJO which acts as a single 
nondegenerate JPA. The microwave routing in our circuit 
which finally combines the two separate nondegenerate
JPAs is explained in the main text.

The JJO is characterized by the 
nonlinear Josephson inductance $L_{J}(\Phi) = h/(4\pi e I_{c}(\Phi))$ and by 
the shunting capacitor $C_{0}$ (compare with 
Fig.~\ref{fig01}(a) of the main text). One can further specify 
the JJO through its characteristic admittance 
$Y_{0} = \sqrt{C_{0}/L_{J}(\Phi)}$, fundamental resonance frequency 
$\omega_{0} = 1/\sqrt{L_{J}(\Phi) C_0}$ and quality factor 
$Q \approx \omega_{0} \mathrm{Re}\left[Z_{in}''''\right] C_0$.
Here, it is as usual $Y_{in}'''' = (Z_{in}'''')^{-1}$. $L_{J}(\Phi)$ is tunable through the
external magnetic flux $\Phi$ piercing through the SQUID loop. 
Since $C_0$ is much larger than the intrinsic 
capacitance of the Josephson junctions, $C_J$, we neglect the latter 
capacitance in the following. Note that here 
$I_{c}(\Phi) = 2i_{c}\left\lvert\cos\left(\pi \frac{\Phi}{\phi_0}\right)\right\rvert$ is the 
total critical current of the Josephson junctions in the SQUID, 
$i_{c}$ denotes the critical current of each of the two Josephson junctions in the 
SQUID and $\phi_0 = h/(2e)$ is the flux quantum (not to be confused with the symbol 
for the node flux of a transmission line we will introduce later). 
Therefore, we assume that the two junctions are identical. The case of non-identical 
junctions is easily incorporated by considering a SQUID asymmetry due to 
which one cannot adjust anymore a perfectly zero critical current. We also 
assume a negligible inductance of the SQUID loop arms which has to be 
compared to the total Josephson inductance.
On the other hand, for small loop inductance $L_{loop}$, 
but still sizable against $L(\Phi)$, 
the minimal reachable critical current in the SQUID increases as 
$\pi L_{loop} I_c/\phi_0$ but is in most of the practical cases still 
small enough to reach a large enough Josephson inductance or 
equivalently to reach a small enough plasma frequency $\omega_{0}$. 

Like in the main text, $\tilde{\omega}_{s} = {\omega}_{p} - {\omega}_{s}$ 
denotes the detuning of the signal from the pump frequency and 
$\tilde{\omega}_{p} = {\omega}_0 - {\omega}_p$ is the detuning of the pump 
frequency from the plasma frequency $\omega_0$ of the JJO. 
For not too large detuning $\tilde{\omega}_{s}$ and $\tilde{\omega}_p$  
from the center (design) frequency of the 
EM environment of 6~GHz, $Y_{in}''''\approx (30.0~\Omega)^{-1}$ 
is to good approximation real valued. Furthermore, 
$Y_{0} \approx (5.45~\Omega)^{-1}$ for the parameter regime envisioned 
in this work for a tunable JJO frequency of 
$\omega_{0} = 1/\sqrt{L_{J}(\Phi)C_{0}} \approx 2\pi\cdot 7.3$~GHz, 
where $C_{0} = 4.0$~pF is fixed through its parallel plate geometry in the 
microstrip circuitry and $L_{J} = 0.12$~nH in this case 
but still magnetic-flux tunable as said before. With this (arbitrary) choice 
of $\omega_0$ we anticipate the yet to be derived result 
(compare with Eq.~(\ref{eq:S23})) that the plasma frequency 
shifts with increasing pumping amplitude.

For the purpose of our paper, 
the dynamics of the phase difference 
$\delta \propto 2\pi\Phi/\phi_0~(\mathrm{mod}~2\pi)$ of the JJO 
is well described by the resistively and capacitively shunted 
Josephson junction (RCSJ) model [M. Tinkham, \emph{Introduction 
to Superconductivity}, Dover Publications, Inc., 2nd edition, 2004]. 
The solution of the RCSJ model yields the intra-oscillator field 
both for the strong pump tone and for the noise (or the weak signal) which are leaking 
via the transmission line into the JJO (cf.~Fig.~\ref{figS02}).
We want to separate $\delta$ into two parts. 
A classical part, $\delta_{p}$, belonging 
to the strong pump and a quantum part, $\hat{\delta}_s$, belonging to the noise 
on the transmission line or/and to a weak signal which is amplified by the device.

Although the RCSJ model is originally formulated for a single Josephson 
junction with only one critical current $i_{c}$, it can be readily translated 
to apply to a Josephson junction SQUID consisting of two junctions in which the critical 
current is magnetic-flux tunable as described before. We, therefore, substitute 
$I_{c}(\Phi)$ in the RCSJ model and treat it as the (total) critical current of the 
system, yet adjustable. Hence, the superconducting response of the 
Josephson junction SQUID is 
covered by a term $I_{c}(\Phi)\sin(\delta(t))$ with the phase difference $\delta$ 
defined before and $I_c$ being now a magnetic-flux tunable supercurrent. 
Furthermore, since we want to operate the JJO 
in the zero-voltage state, we assume that it is only shunted by the 
radio-frequency (\emph{rf}) resistance of its EM environment and not by 
its normal-state resistance which would otherwise appear in the 
voltage-carrying state of the junction. In this particular situation, 
the RCSJ model goes over into the Duffing equation after Taylor expansion of 
the '$\sin(\delta(t))$' term in the RCSJ model 
and keeping only the first non-linear term. 
The Duffing equation finally reads:
\begin{equation}
\label{eq:S01}
\frac{C_0 \phi_0}{2\pi} \ddot\delta(t) + \frac{Y''''_{in} \phi_0}{2\pi} \dot \delta(t) 
+ I_{c}(\Phi)\left[\delta(t) - \frac{1}{6}\delta(t)^3\right] = I(t) ~,
\end{equation}
where $C_0$ is the shunting capacitance 
of the JJO and $Y''''_{in}$ is the input admittance of the EM environment at the pumping frequency.
Furthermore, $I(t) = I_{p}(t) + \hat{I}_{s}(t)$ is the total microwave pump current 
applied externally to the JJO, 
consisting of the strong pump current $I_{p}$ which 
is treated in the following as a classical signal. 
Additionally, the JJO is also pumped by the quantum component $\hat{I}_s$ 
via the same port like the pump current.

Dividing Eq.~(\ref{eq:S01}) by the term '$C_0\phi_0/(2\pi)$', the Duffing 
equation assumes a form in which the characteristic parameters of the 
JJO explicitly appear:
\begin{equation}
\label{eq:S02}
\ddot\delta(t) + \frac{Y''''_{in}}{C_0} \dot \delta(t) 
+ \omega_{0}^{2}\left[\delta(t) - \frac{1}{6}\delta(t)^3\right] = \omega_{0}^{2} \frac{I(t)}{I_{c}(\Phi)} ~,
\end{equation}
where we identify the plasma frequency of the JJO as 
$\omega_{0} = \sqrt{\frac{2\pi I_{c}(\Phi)}{\phi_0 C_0}}$ or equivalently written as 
$\omega_{0} = 1/\sqrt{L_{J}(\Phi) C_0}$ as stated already before.

In order to prepare the evaluation of the intra-oscillator field $\delta$ of 
the pumped JJO we proceed in two steps.
First, we are interested in the 
large-signal solution of Eq.~(\ref{eq:S02}), 
very similar to the solution strategy in quantum SIS heterodyne mixer theory 
[J.~R.~Tucker and M.~J.~Feldman, Rev.~Mod.~Phys.~{\bf 57}, 1055, 1985] when 
predicting the current-voltage characteristic, gain and noise of the mixer device.
In a second step described further below, we rewrite Eq.~(\ref{eq:S02}) to account for 
a weak quantum signal which is coupled to the oscillator; this form of the equation is 
used to calculate the parametric gain and noise of the JPA in the main text of our work.  

For the first step, we assume that no quantum signal is coupled into the oscillator, 
$\hat{I}_{s}(t) = 0$, and only the strong pump current 
$I_{p}(t) = \bar{I}_{p} \cos(\omega_{p} t )$ is applied pumping the 
nonlinear JJO. In this situation the system behaves 
like a classical nonlinear oscillator and we do not have to further specify 
the complications which have to be introduced when describing the system 
quantum mechanically (like we will do further below after the following 
brief preparation).

In the method of harmonic balance one assumes 
a solution of the Duffing equation (\ref{eq:S02}) of the form
$\delta_{p} = a\cos(\omega_{p}t) + b\sin(\omega_{p}t)$ or equivalently 
$\delta_{p} = \delta_{p,max} \cos(\omega_{p} t -\varphi)$ where 
$\delta_{p,max}^{2} = a^2 + b^2$ and $\tan(\varphi) = b/a$. Therefore, 
we separate the solution into an in-phase ($\propto \cos(\cdot)$) and a
quadrature phase ($\propto \sin(\cdot)$) term (I\&Q term). After substituting 
$\delta_{p}$ into (\ref{eq:S02}) 
and combining/simplifying the resulting terms one arrives 
at the following equation system:
\begin{widetext}
\begin{equation}
\label{eq:S03}
\begin{split}
&\cos(\omega_{p}t) \left[-a\omega_{p}^{2} + \frac{b\omega_{p} Y''''_{in}}{C_0}
+ a\omega_{0}^{2} - \frac{1}{6}\omega_{0}^{2}
\left(\frac{3a^{3}}{4}+\frac{3ab^{2}}{4}\right) 
- \omega_{0}^{2} \frac{\bar{I}_{p}}{I_{c}}\right]\\
+&\sin(\omega_{p}t)\left[-b\omega_{p}^{2} - \frac{a\omega_{p} Y''''_{in}}{C_0}
+b\omega_{0}^{2} - \frac{1}{6}\omega_{0}^{2}
\left(\frac{3b^{3}}{4}+\frac{3a^{2}b}{4}\right)\right]\\
+&\cos(3\omega_{p}t)\left[-\frac{1}{6}\omega_{0}^{2}
\left(\frac{a^3}{4}-\frac{3ab^2}{4}\right)\right]\\
+&\sin(3\omega_{p}t)\left[-\frac{1}{6}\omega_{0}^{2} 
\left(-\frac{b^3}{4}+\frac{3a^{2}b}{4}\right)\right] = 0~.
\end{split}
\end{equation}
\end{widetext}
In the following we will neglect the higher order harmonics 
$\propto \cos(3\omega_{p} t)$ and $\propto \sin(3\omega_{p} t)$. 
Furthermore, the $\cos(\cdot)$ and $\sin(\cdot)$ 
terms can be seen as an orthogonal function basis so that we can 
proceed as follows. Squaring the first and second line in 
Eq.~(\ref{eq:S03}) and adding the results 
leads to the large signal solution of the Duffing equation which 
after further simplifications assumes the following form:
\begin{figure}[tb]
\centering
\includegraphics[width=\columnwidth]{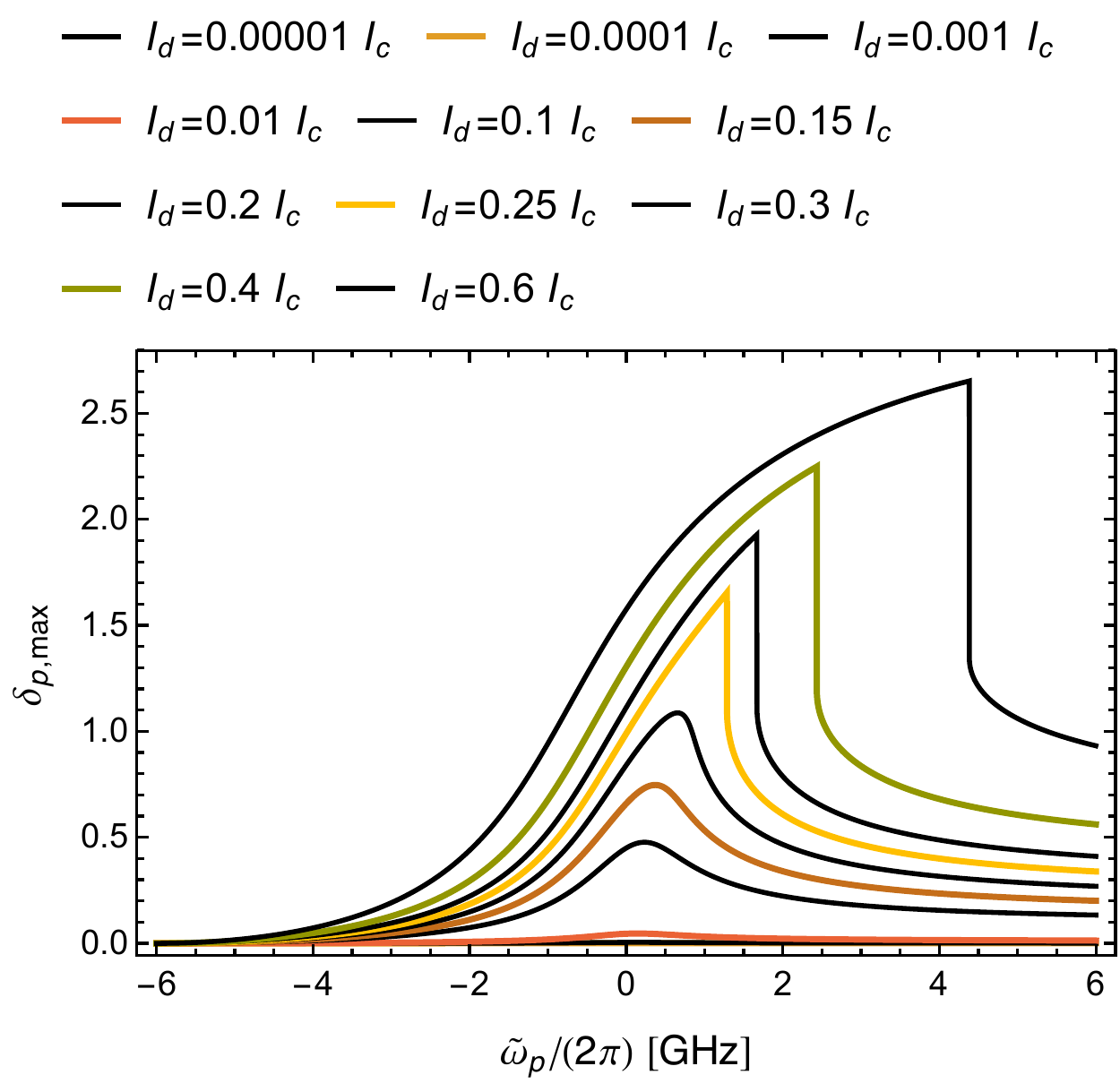}
\caption{\label{figS01}Nonlinear frequency response of the classically 
pumped JJO in the large signal picture, 
evaluated by solving Eq.~(\ref{eq:S04}) for $\delta_{p,max}$. 
The rms pumping amplitude is expressed in terms of the critical current $I_c$
of the Josephson junction SQUID. Parameters used for the plot: $C_0 = 4$~pF and 
$(Y''''_{in}[\omega_{p}])^{-1} = 30~\Omega$ for a constant pump frequency of 
$\omega_{p}/(2\pi) = 6~\mathrm{GHz}$. The x-axis 
shows the pump frequency detuning from the plasma frequency 
$\tilde\omega_{p}/(2\pi) = (1/2\pi)(\omega_{0} - \omega_{p})$ 
and the y-axis shows the internal field in the oscillator building up
due to the pumping. With increasing pumping amplitude, the maximum 
response shifts to increasingly positive pump frequency detuning. The optimal 
operating point of the amplifier is where a small signal change induces a big 
change in the oscillator dynamics at a detuning of approximately 
$\tilde\omega_{p}/(2\pi) = 1.5~\mathrm{GHz}$ or in other words at plasma frequencies 
of about $\omega_{0} = 7.3 \textrm{-} 7.5~\mathrm{GHz}$ like found consistently 
in the main text.}
\end{figure}
\begin{equation}
\label{eq:S04}
\begin{split}
&\left[\left(\omega_{p}^{2} - \omega_{0}^{2} + 
\frac{\omega_{0}^{2}\delta_{p,max}^{2}}{8}\right)^{2} + 
\left(\frac{\omega_{p}Y''''_{in}}{C_0}\right)^{2} \right]\delta_{p,max}^{2}\\
&= \omega_{0}^{4} \frac{\bar{I}_{p}^{2}}{I_{c}^{2}(\Phi)}~.
\end{split}
\end{equation}
Equation~(\ref{eq:S04}) describes the full nonlinear frequency response of the 
JJO in the large-signal picture which we show for different 
pumping strengths in Fig.~\ref{figS01}.
For increasing pumping strength, the maximum 
oscillator frequency response shifts to increasingly large pump 
frequency detuning and becomes nonlinear as expected from Eq.~(\ref{eq:S04}). 
This has to be taken into consideration when 
adjusting the operation frequency of the JPA via the 
magnetic-flux, tuning $\omega_{0}$.
\begin{figure}[tb]
\centering
\includegraphics[width=\columnwidth]{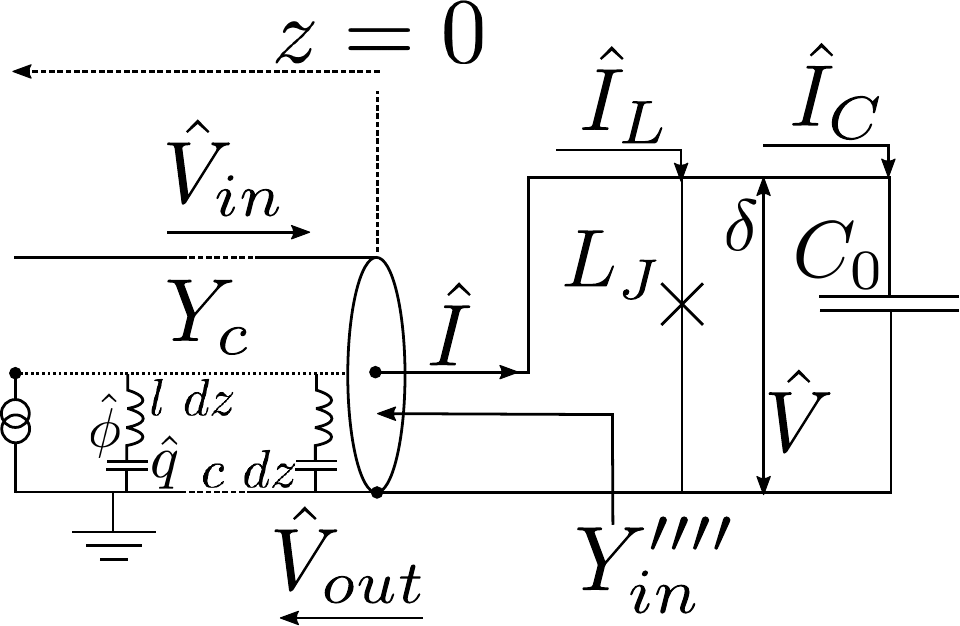}
\caption{\label{figS02}
{\bf Input, output and intra-oscillator ($\delta$) fields in the nonlinear 
JJO}.
The nonlinear oscillator is pumped by a microwave current source and 
consists of a nonlinear inductor (Josephson junction or SQUID; $\times$ symbol) 
and linear capacitor connected in parallel to a transmission line of characteristic 
admittance $Y_c$, providing the input admittance $Y_{in}''''$ 
(compare with Fig.~\ref{fig01}(a) of the main text) to the 
nonlinear oscillator at $z=0$.
This admittance is the only source of dissipation in the system which
allows for exchange of energy into and out of the nonlinear oscillator with rate $\kappa$. 
The strong pump and the weak (quantum) 
noise and signal field leak into the nonlinear oscillator and subsequently 
the internal oscillator field leaks out again into the 
transmission line where it is connected to the measurement 
apparatus. The nonlinear inductance does provide parametric 
amplification of the weak input field when it is additionally 
pumped by the strong pumping field. The conjugated quantum 
variables of the nonlinear oscillator are the phase and 
charge across the nonlinear inductor and linear capacitor.}
\end{figure}

In the second step we aim at understanding the nonlinear oscillator 
dynamics and, hence, how the parametric amplification is established 
when we couple along with the strong pump tone, a weak 
signal or (quantum) noise into the oscillator. With quantum noise we 
specifically mean the vacuum fluctuations on the 
transmission line which are 
injected into the JPA. We will specify in the next section of this 
supplemental information what we exactly mean with that and why 
it is important to discuss this topic when dealing 
with (quantum) amplification with a phase-preserving, 
i.e.~nondegenerate JPA.

Our following task will be to derive how Eq.~(\ref{eq:S02})
is modified when we couple a weak quantum signal to the nonlinear 
oscillator; a sketch of this situation is shown in Fig.~\ref{figS02}.

To do so, first, we have to describe our circuit quantum mechanically. 
We will then come back to the derivation of the nonlinear oscillator equation 
describing pumping through a weak quantum signal further below.
 
We recall that to perform the circuit quantization, 
one describes a (quantum) transmission line or an oscillator as a set of 
noninteracting bosonic modes like shown on the positive $z$-side 
in Fig.~\ref{figS02}, suggesting 
an infinite chain of inductors and shunting capacitors in the inside of the 
transmission line leading to an 
admittance $Y_c$ [A.~O.~Caldeira and A.~J.~Leggett, 
Ann.~Phys.~(N.~Y.)~{\bf 149}, 374, (1983)]. 
This description is well established and is based on 
the Lagrangian/Hamiltonian formulation of the transmission line and 
nonlinear oscillator to which the transmission line 
is connected to (see for an instructive 
review [A.~A.~Clerk et al., Rev.~Mod.~Phys.~{\bf 82}, 
1155, (2010)] and by considering in particular also [M.~ Devoret, 
\emph{Quantum Fluctuations in Electrical Circuits}, 
Les Houches, Sessions LXIII, (1995)] and [B.~Yurke and 
J.~S.~Denker, Phys.~Rev.~A.~{\bf 29}, 1419, (1984)]). 
Quantization is achieved by elevating the node flux $\phi$ and charge $q$ of the 
transmission line to quantum operators which generate the 
traveling fields. Below we bring into context
the main results of such a description for the purposes in our paper and refer
for further details to the references mentioned before.

The Hamiltonian of the transmission line reads: 
\begin{equation}
\label{eq:S05}
\hat{H} = \int dz \left[\frac{\hat{q}^2(z,t)}{2c} + 
\frac{1}{2l}\left(\frac{\partial}{\partial z}\hat{\phi}(z,t)\right)^2\right]~.
\end{equation}
The expression in square brackets is the Hamiltonian density 
of some element $dz$ in the transmission line like suggested in 
Fig.~\ref{figS02} and the total Hamiltonian of the element is obtained by
summing over all elements $dz$, yielding essentially the energy density 
in the transmission line, evaluated by the integral. Flux $\hat{\phi}$ 
across the inductor and charge density $\hat{q}$ on the capacitor 
are conjugated variables and yield the voltages and currents on the
transmission line; $\hat{\phi}(z,t) = \int_{-\infty}^{t} d\tau \hat{V}(z,\tau)$,  
$\hat{I}(z,t) = -\frac{1}{l} \frac{\partial \hat{\phi}(z,t)}{\partial z}$ and 
$c\frac{\partial\hat{\phi}}{\partial t} = c \hat{V}(z,t) = \hat{q}(z,t)$ with 
$c$ and $l$ being the capacitance and inductance per unit line length.
Quantum mechanics enters the description of the transmission line 
by identifying \emph{equal-time} commutation relations for the conjugated
variables, $\left[\hat{\phi}(z,t),\hat{q}(z',t)\right] = i\hbar\delta(z-z')$ 
and $\left[\hat{q}(z',t),\hat{q}(z,t)\right] = \left[\hat{\phi}(z',t),\hat{\phi}(z,t)\right] = 0$ 
where we have taken the limit $dz\rightarrow 0$.

From the Hamiltonian equation of motion (or alternatively from the 
Euler-Lagrange equation) for the flux variable one finds 
that the flux obeys a 1D wave equation; 
$\frac{\partial^{2}\hat{\phi}(z,t)}{\partial z^{2}} - 
\frac{1}{v_{ph}} \frac{\partial^{2}\hat{\phi}(z,t)}{\partial z^{2}} = 0$ 
where $v_{ph} = \frac{1}{\sqrt{l c}}$ is the phase velocity. 
This form of wave equation can be interpreted as being a 
massless Klein-Gordon equation with the photon (excitation) on the 
transmission line being the fundamental particle.

The solution of the wave equation can be decomposed into an ingoing ('in') and 
an outgoing ('out') component where in Fig.~\ref{figS02} the 'in' part travels from positive
$z$-coordinates towards '0' where the nonlinear oscillator is connected, and the 'out' part travels 
in the opposite direction towards the rest of the circuit and finally to the readout of the amplifier: 
\begin{equation}
\begin{split}
\label{eq:S06}
\hat{\phi}_{in}(z,t) &= \sqrt{\frac{\hbar}{2 Y_{c}}} \int_{0}^{\infty} \frac{d\omega}{2\pi\sqrt{\omega}}
\left(i\hat{a}_{in}[\omega]e^{-i\omega(t-\frac{z}{v_{ph}})} + \mathrm{h.c.}\right)\\
\hat{\phi}_{out}(z,t) &= \sqrt{\frac{\hbar}{2 Y_{c}}} \int_{0}^{\infty} \frac{d\omega}{2\pi\sqrt{\omega}}
\left(i\hat{a}_{out}[\omega]e^{-i\omega(t+\frac{z}{v_{ph}})} + \mathrm{h.c.}\right)~.
\end{split}
\end{equation}
In total $\hat{\phi} = \hat{\phi}_{in} + \hat{\phi}_{out}$ and 
'\emph{h.c.}' stands for the hermitian conjugate of the term in brackets. Note that the 
characteristic admittance can be written as $Y_{c} = \sqrt{c/l}$ and 
that there is a certain arbitrariness in the choice of the phase 
(yielding an extra '$i$' and in certain cases also an extra sign in front of 
the equations) which has to be chosen in such 
a way to fulfill the operator commutation relations and to yield a consistent 
voltage on the transmission line. We will discuss this shortly.

In a next step one can
formulate a quantized voltage. The in- and outgoing quantized 
voltages in the circuit at the point $z = 0$
then read (c.f.~Fig.~\ref{figS02}):
\begin{equation}
\begin{split}
\label{eq:S07}
\hat{V}_{in}(t) &= \sqrt{\frac{\hbar}{2Y_{c}}}\int_{0}^{\infty} 
\frac{d\omega}{2\pi}\sqrt{\omega} \left(\hat{a}_{in}[\omega]e^{-i\omega t} 
+ \mathrm{h.c.}\right)
\\
\hat{V}_{out}(t) &= \sqrt{\frac{\hbar}{2Y_{c}}}\int_{0}^{\infty} 
\frac{d\omega}{2\pi}\sqrt{\omega} \left(\hat{a}_{out}[\omega]e^{-i\omega t}
+ \mathrm{h.c.}\right)~,
\end{split}
\end{equation}
and as before for the phase, also for the voltage $\hat{V} = \hat{V}_{in} + \hat{V}_{out}$.
The charge density operator $\hat{q}_{in,out}$ on the transmission line 
is obtained by multiplying Eq.~(\ref{eq:S07}) by the capacitance 
per unit line length $c$. Note that the creation and annihilation operators 
of the 'in' fields in frequency domain ($\hat{a}_{in}^{\dagger}$ and 
$\hat{a}_{in}$) have a dimension of $\sqrt{s}$. Furthermore, they 
obey the following commutation relations (written below only for the 'in' operators):
\begin{equation}
\begin{split}
\label{eq:S08}
\left[\hat{a}_{in}[\omega],\hat{a}_{in}[\omega']\right]&=
\left[\hat{a}^{\dagger}_{in}[\omega],\hat{a}^{\dagger}_{in}[\omega']\right] = 0,~\mathrm{and}\\
\left[\hat{a}_{in}[\omega],\hat{a}^{\dagger}_{in}[\omega']\right]&=2\pi\delta(\omega-\omega')~.
\end{split}
\end{equation}
The 'in' and 'out' fields do always commute with each other 
(for instance $\left[\hat{a}_{in}[\omega],\hat{a}^{\dagger}_{out}[\omega']\right] = 0$).
As a quick sanity check, we verify whether the results 
Eq.~(\ref{eq:S06})-(\ref{eq:S08}) fulfill the commutation relations
for the conjugated variables $\hat{\phi}$ and $\hat{q}$.
The commutator for the flux and charge density reads:
\pagebreak
\begin{widetext}
\begin{equation}
\begin{split}
\label{eq:S09}
\Bigl[\hat{\phi}(z,t),\hat{q}(z',t)\Bigr]
=&\hat{\phi}_{in}(z,t)\hat{q}_{in}(z',t) + \hat{\phi}_{in}(z,t)\hat{q}_{out}(z',t)
+\hat{\phi}_{out}(z,t)\hat{q}_{in}(z',t) + \hat{\phi}_{out}(z,t)\hat{q}_{out}(z',t)\\
&-\hat{q}_{in}(z',t)\hat{\phi}_{in}(z,t) - \hat{q}_{in}(z',t)\hat{\phi}_{out}(z,t) 
- \hat{q}_{out}(z',t)\hat{\phi}_{in}(z,t) -  \hat{q}_{out}(z',t)\hat{\phi}_{out}(z,t)~,
\end{split}
\end{equation}
\end{widetext}
where the mixed 'in/out' operator products are zero according to 
the commutation relations for the fields, Eq.~(\ref{eq:S08}).
Finally, we are left with the following expression which consists of 
four main terms (i)-(iv):
\begin{widetext}
\begin{equation}
\label{eq:S10}
\underbrace{\hat{\phi}_{in}(z,t)\hat{q}_{in}(z',t)}_{\mathrm{(i)}}
+ \underbrace{\hat{\phi}_{out}(z,t)\hat{q}_{out}(z',t)}_{\mathrm{(ii)}}
-\underbrace{\hat{q}_{in}(z',t)\hat{\phi}_{in}(z,t)}_{\mathrm{(iii)}}
-\underbrace{\hat{q}_{out}(z',t)\hat{\phi}_{out}(z,t)}_{\mathrm{(iv)}}~.
\end{equation}
\end{widetext}
The difference between term (i) and term (iii) 
yields half the value of the commutator in Eq.~(\ref{eq:S09}):
\begin{widetext}
\begin{equation}
\begin{split}
\label{eq:S11}
&\hat{\phi}_{in}(z,t)\hat{q}_{in}(z',t) - \hat{q}_{in}(z',t)\hat{\phi}_{in}(z,t)=
c\frac{i\hbar}{8\pi^{2}Y_{c}} \int_{\mathcal{D}}d\omega' d\omega
\Biggl\lbrace\Biggr.\left(\hat{a}_{in}[\omega]e^{-i\omega\left(t-\frac{z}{v_{ph}}\right)}
\hat{a}^{\dagger}_{in}[\omega']e^{+i\omega'\left(t-\frac{z'}{v_{ph}}\right)}\right)\\
&- \left(\hat{a}^{\dagger}_{in}[\omega]e^{+i\omega\left(t-\frac{z}{v_{ph}}\right)}
\hat{a}_{in}[\omega']e^{-i\omega'\left(t-\frac{z'}{v_{ph}}\right)}\right) 
+\left(\hat{a}_{in}[\omega']e^{-i\omega'\left(t-\frac{z'}{v_{ph}}\right)}
\hat{a}^{\dagger}_{in}[\omega]e^{+i\omega\left(t-\frac{z}{v_{ph}}\right)}\right)\\
&-\left(\hat{a}^{\dagger}_{in}[\omega']e^{+i\omega'\left(t-\frac{z'}{v_{ph}}\right)}
\hat{a}_{in}[\omega]e^{-i\omega\left(t-\frac{z}{v_{ph}}\right)}\right)
\Biggl.\Biggr\rbrace\\
&=c\frac{i\hbar}{8\pi^{2}Y_{c}}\int_{\mathcal{D}}d\omega'd\omega 
\Biggl\lbrace\Biggl[\hat{a}_{in}[\omega],\hat{a}_{in}^{\dagger}[\omega']\Biggr]
e^{-i\left(\omega - \omega'\right)t - i\left(\omega' \frac{z'}{v_{ph}} - \omega\frac{z}{v_{ph}}\right)}
+ \Biggl[\hat{a}_{in}[\omega'],\hat{a}_{in}^{\dagger}[\omega]\Biggr]
e^{-i\left(\omega' - \omega\right)t - i\left(\omega \frac{z}{v_{ph}} - \omega'\frac{z'}{v_{ph}}\right)}
\Biggr\rbrace\\
&=c\frac{i\hbar}{8\pi^{2}Y_{c}}\int_{\mathcal{D}}d\omega'd\omega 
\Biggl\lbrace2\pi\delta(\omega-\omega')
e^{-i\left(\omega - \omega'\right)t - i\left(\omega' \frac{z'}{v_{ph}} - \omega\frac{z}{v_{ph}}\right)}
+ 2\pi\delta(\omega'-\omega) 
e^{-i\left(\omega' - \omega\right)t - i\left(\omega \frac{z}{v_{ph}} - \omega'\frac{z'}{v_{ph}}\right)}
\Biggr\rbrace\\
&=\frac{i\hbar}{4\pi v_{ph}}\int_{0^{-}}^{\infty}d\omega 
\Biggl\lbrace e^{- i\frac{\omega}{v_{ph}}\left(z' - z\right)}
+e^{- i\frac{\omega}{v_{ph}}\left(z - z'\right)}
\Biggr\rbrace
=
\frac{i\hbar}{2}\int_{0^{-}}^{\infty}\frac{dk}{2\pi} 
\Biggl\lbrace e^{- ik\left(z' - z\right)}
+e^{- ik\left(z - z'\right)}\Biggr\rbrace = \frac{i\hbar}{2}\delta(z-z')~,
\end{split}
\end{equation}
\end{widetext}
where in the last steps we have expressed $\omega = v_{ph} k$ with $k$ 
being the wave vector. Also, we have already taken into account that 
equal pairs of creation and annihilation operators 
for the fields are equal to zero (c.f.~Eq.~(\ref{eq:S08})) and the integration contour
$\mathcal{D}$ is for now the positive $\mathbb{R}^{2}$. 

We obtain the same 
result like in Eq.~(\ref{eq:S11}) for the difference between the terms (ii) and (iv)
in Eq.~(\ref{eq:S10}) and have, therefore, verified that the commutation 
relation for the flux and charge density are correct and, consequently, also 
the relations Eq.~(\ref{eq:S06}) and Eq.~(\ref{eq:S07}).

The equal pair commutation relations $\left[\hat{\phi}(z,t),\hat{\phi}(z',t)\right] = 0$ and 
$\left[\hat{q}(z,t),\hat{q}(z',t)\right] = 0$ are trivially fulfilled which is obtained by 
a similar calculation like the one before.

We further assume the following Fourier relations for the annihilation and creation operators: 
\begin{equation}
\begin{split}
\label{eq:S12}
\hat{a}_{in}(t) &= \int_{-\infty}^{\infty}\frac{d\omega}{2\pi}~\hat{a}_{in}[\omega]e^{-i\omega t}\\
\hat{a}_{in}[\omega] &= \int_{-\infty}^{\infty}dt~\hat{a}_{in}(t)e^{+i\omega t}\\
\hat{a}_{in}^{\dagger}(t) &= \int_{-\infty}^{\infty}\frac{d\omega}{2\pi}~\hat{a}^{\dagger}_{in}[\omega]e^{-i\omega t}\\
\hat{a}_{in}^{\dagger}[\omega] &= \int_{-\infty}^{\infty}dt~\hat{a}_{in}^{\dagger}(t)e^{+i\omega t}~,
\end{split}
\end{equation}
and correspondingly for the 'out' fields, following the convention suggested in
[A.~A.~Clerk et al., Rev.~Mod.~Phys.~{\bf 82}, 1155, 2010]. Note that in this 
convention the sign of the exponentials of the time or frequency 
$\hat{a}$ and $\hat{a}^{\dagger}$ 
operators is the same in contrast to the definition which is usually used in 
quantum optics where the signs are opposite. 
The practical reason for our definition of these operators is that one can 
express now two different frequency modes (for us this is the signal and 
idler mode of the JPA) by specifying either $\hat{a}[\omega]$ (signal) or 
$\left(\hat{a}[-\omega]\right)^{\dagger} = \hat{a}^{\dagger}[\omega]$ (idler) as 
described by [A.~A.~Clerk et al., Rev.~Mod.~Phys.~{\bf 82}, 1155, 2010].

With the definitions Eq.~(\ref{eq:S12}) one obviously transforms also the 
fields $\hat{V}$, $\hat{\phi}$ and $\hat{q}$ 
from time into frequency domain and back with the same sign convention 
as discussed before.

Sometimes it is easier to interpret fields of the form like in Eq.~(\ref{eq:S07}) by 
going over to the Markov approximation in which one assumes 
a sharp enough frequency response of the nonlinear oscillator around its plasma 
frequency $\omega_{0}$. One obtains then:
\begin{equation}
\label{eq:S13}
\hat{V}_{in}(t) = \sqrt{\frac{\hbar \omega_{0}}{2Y_{c}}}
\left(\hat{a}_{in}(t)e^{-i\omega_{0} t} 
+ \hat{a}^{\dagger}_{in}(t)e^{+i\omega_{0} t}\right)~,
\end{equation}
and corresponding equations for the other fields.
In this case, however, 
the interpretation of the 
creation and annihilation fields is different and they represent now 
slow varying envelopes compared to the 
oscillation frequency $\omega_{0}$ of the nonlinear oscillator.

With this preparation at hand, 
a spatially dependent admittance on a transmission line can then be defined as:
\begin{equation}
\label{eq:S14}
Y[z,\omega] = \frac{I[z,\omega]}{V[z,\omega]} = \left(Z[z,\omega]\right)^{-1}~,
\end{equation}
which allows in turn to define a reflection coefficient (in essence the ratio between 
in- and output fields):
\begin{equation}
\label{eq:S15}
r[z,\omega] = \frac{Z[z,\omega] - Z_{c}}{Z[z,\omega] - Z_{c}}~,
\end{equation}
with $Z_{c} = Y_{c}^{-1}$ being the characteristic impedance of the transmission line on which 
the reflection coefficient is measured. The reflection 
coefficient can obviously also be defined as $V_{out} = r V_{in}$.  
By the choice of the symbol '$r$' instead of '$\Gamma$' 
which is usually used in classical microwave theory to express the reflection coefficient, 
we choose a formalism used to describe parametric amplifiers in the input-output formalism.

Having a language defined how we can describe a circuit quantum mechanically, we 
come back to formulate a nonlinear oscillator equation describing 
pumping through a weak quantum signal send into the oscillator via the (quantum) transmission 
line which is connected to it.

The current in the circuit $\hat{I}(z,t) = Y_{c}\left(\hat{V}_{in}(z,t) - \hat{V}_{out}(z,t)\right)$, 
shown in Fig.~\ref{figS02}, has to obey \emph{Kirchhoff's} law and we will focus now on the 
point $z = 0$ which is the point at which the transmission line connects to the nonlinear
oscillator:
\begin{equation}
\begin{split}
\label{eq:S16}
&-\hat{I}(t) + \hat{I}_{L}(t) + \hat{I}_{C}(t) = 0\\
&\Rightarrow\hat{I}_{L}(t) + Y_{in}''''\hat{V}(t) + \hat{I}_{C}(t) = 2Y_{in}''''\hat{V}_{in}(t)~,
\end{split}
\end{equation}
where we have written $\hat{V}_{out}(t) = \hat{V}(t) - \hat{V}_{in}(t)$ and explicitly 
identified the input admittance $Y_{in}''''$ at the point $z=0$ of the transmission line in consistence 
to the notation we use in the main text. At the same time this admittance, however, 
is to good approximation equal to the characteristic admittance of the transmission line over 
the operation bandwidth of the parametric amplifier but leaves space for a more general description of the 
circuit since it carries a real and a complex part which is tunable by our circuit design.
The right side of Eq.~(\ref{eq:S16}) is the pumping term of the nonlinear oscillator due to 
a weak (quantum signal). Going back to Eq.~(\ref{eq:S02}) we can write the total 
pumping term $I(t)$ then as:
\begin{equation}
\label{eq:S17}
I(t) = \bar{I}_{p} \cos(\omega_{p} t ) + 2Y_{in}''''\hat{V}_{in}(t) = 
\bar{I}_{p} \cos(\omega_{p} t ) + 2\hat{I}_{in}(t)~,
\end{equation}
consisting of a classical pump tone and the weak quantum 
signal ($\hat{I}_{in}$). Therefore, the equation describing the nonlinear 
oscillator, pumped by a classical source and a quantum signal can be written as: 
\begin{equation}
\begin{split}
\label{eq:S18}
\ddot\delta(t) + \frac{Y''''_{in}}{C_0} \dot \delta(t) 
+ \omega_{0}^{2}\left[\delta(t) - \frac{1}{6}\delta(t)^3\right] 
&- \omega_{0}^{2} \frac{\bar{I}_{p} \cos(\omega_{p} t)}{I_{c}(\Phi)}\\
&= \frac{4\pi \hat{I}_{in}(t)}{\phi_{0}C_{0}}~.
\end{split}
\end{equation}
Note that the phase $\delta$ contains now the two 
components introduced already at the beginning of this supplemental information, 
the classical component $\delta_{p}$ which we 
summarize for different pumping strengths in Fig.~\ref{figS01} and the 
quantum (or weak signal) component $\hat{\delta}_{s}$ which will be the subject of 
our further discussion and the central variable to calculate the parametric gain.

Since $\hat{\delta}_{s} \ll \delta_{p}$, it 
can be treated as a small perturbation around the solution of the strong pump 
$\delta_{p}$. To first order the solution reads then 
$\delta= \delta_{p} + \hat{\delta}_{s}$ which is nothing more than a first order 
perturbation expansion of $\hat{\delta}_{s}$ around the large-signal solution of the 
nonlinear oscillator which we have already solved before.

We substitute $\delta(t) = \delta_{p}(t) + \hat{\delta}_{s}(t)$ into Eq.~(\ref{eq:S18}) and retain 
only linear terms in $\hat{\delta}_{s}(t)$. Furthermore, we subtract the large signal equation 
of motion (i.e.~the equation of motion for the classical phase $\delta_{p}$) from the equation 
of motion of $\delta$ (Eq.~(\ref{eq:S18})) in order to obtain the equation of motion 
describing the dynamics of the intra-oscillator quantum variable $\hat{\delta}_{s}(t)$. 
We finally obtain:
\begin{widetext}
\begin{equation}
\label{eq:S19}
\ddot{\hat{\delta}}_{s}(t) + \frac{Y''''_{in}}{C_0} \dot{\hat{\delta}}_{s}(t)
+ \omega_{0}^{2} \left\lbrace 1 - \frac{\delta_{p,max}^{2}}{4}
\left[\cos\left(2\omega_{p}t - 2\varphi \right) + 1\right]\right\rbrace\hat{\delta}_{s}(t) 
= \frac{4\pi \hat{I}_{in}(t)}{\phi_{0}C_{0}} = \frac{4\pi Y''''_{in}\hat{V}_{in}(t)}{\phi_{0}C_{0}}~.
\end{equation}
\end{widetext}
In the argument of the $\cos(\cdot)$-term, we recall that the $\varphi$-term is related to the 
in-phase and quadrature terms of the large signal solution, cf.~Eq.~(\ref{eq:S03}).
Also, we write for the intra-oscillator fields:
\begin{equation}
\begin{split}
\label{eq:S20}
\hat{\delta}_{s}(t) &= \int_{0}^{\infty}\frac{d\omega}{2\pi}
\left(\hat{\delta}_{s}[\omega]e^{-i\omega t}
+ \hat{\delta}^{\dagger}_{s}[\omega]e^{+i\omega t}
\right)~\mathrm{and}\\
\hat{\delta}_{s}[\omega] &= \int_{0}^{\infty}dt\left(\hat{\delta}_{s}(t)e^{+i\omega t}
+ \hat{\delta}^{\dagger}_{s}(t)e^{-i\omega t}
\right)
\end{split}
\end{equation}
at the point $z=0$ of the circuit shown in Fig.~\ref{figS02}.
We will show further below how the intra-oscillator field $\hat{\delta}_{s}$ 
is connected to the in- and output fields via the input-output formalism.

Our goal is now to go over to the frequency domain representation of 
Eq.~(\ref{eq:S19}). 
Fourier transformation of Eq.~(\ref{eq:S19}) yields the following equation of motion 
in the frequency domain at the point $z=0$ in the circuit shown in Fig.~\ref{figS02}:
\begin{widetext}
\begin{equation}
\begin{split}
\label{eq:S21}
\left[-\omega^{2} - i\omega\kappa[\omega] 
+ \omega_{0}^{2}\left(1-\frac{\delta_{p,max}^{2}}{4}\right)\right]\hat{\delta}_{s}[\omega]
- \frac{\omega_{0}^{2}\delta_{p,max}^{2}}{8}
e^{+i2\varphi}\hat{\delta}^{\dagger}_{s}\left[-\omega + 2\omega_{p}\right]
&=\frac{4\pi Y''''_{in}[\omega]\hat{V}_{in}[\omega]}{\phi_{0}C_{0}}\\
&=\frac{4\pi \kappa[\omega]}{\phi_{0}}\sqrt{\frac{\hbar \omega}{2Y''''_{in}[\omega]}}
\hat{a}_{in}[\omega]~.
\end{split}
\end{equation}
\end{widetext}
In detail, we use the Fourier convolution theorem to obtain the Fourier transformation 
of the term containing the $\cos(\cdot)$ function in Eq.~(\ref{eq:S19}):
\begin{widetext}
\begin{equation}
\begin{split}
\label{eq:S22}
\mathcal{F}\left[\hat{\delta}_{s}(t)\cdot\cos\left(2\omega_{p}t-2\varphi\right)\right]
&= \mathcal{F}\left[\hat{\delta}_{s}(t)\right] 
\ast \mathcal{F}\left[\cos\left(2\omega_{p}t-2\varphi\right)\right]\\
&=\frac{1}{2}\int_{-\infty}^{\infty}d\omega' 
\hat{\delta}_{s}[\omega']
\left(e^{+i2\varphi}\delta_{Dirac}\left[-2\omega_{p} + \omega-\omega'\right] 
+ e^{-i2\varphi}\delta_{Dirac}\left[+2\omega_{p} + \omega-\omega'\right]\right)\\
&=\frac{1}{2}\left(e^{+i2\varphi}\hat{\delta}_{s}\left[\omega - 2\omega_{p}\right]
+
e^{-i2\varphi}\hat{\delta}_{s}\left[\omega + 2\omega_{p}\right]\right)\\
&=\frac{1}{2}\left(e^{+i2\varphi}\hat{\delta}^{\dagger}_{s}\left[-\omega + 2\omega_{p}\right]
+
e^{-i2\varphi}\hat{\delta}^{\dagger}_{s}\left[-\omega - 2\omega_{p}\right]
\right)\\
&\approx\frac{1}{2}e^{+i2\varphi}\hat{\delta}^{\dagger}_{s}
\left[-\omega + 2\omega_{p}\right]~.
\end{split}
\end{equation}
\end{widetext}
Here, $\mathcal{F}\left[\cdot\right]$ denotes the Fourier transform, $\ast$ is the 
symbol for the convolution product and $\delta_{Dirac}(\cdot)$
is the Dirac distribution. For the transformation 
$\hat{\delta}_{s}\rightarrow\hat{\delta}^{\dagger}_{s}$
we use the definitions (\ref{eq:S12}) for the operators and Eq.~(\ref{eq:S20}).
In order to obtain the last expression in Eq.~(\ref{eq:S22}) we recognize that the 
second term $\propto \hat{\delta}_{s}^{\dagger}[-\omega-2\omega_{p}]$ decays 
much faster than the first term. In Eq.~(\ref{eq:S21}) we identify also 
$\kappa = (Y''''_{in}[\omega]/C_{0})$ 
being the damping rate introduced by the nonlinear oscillator. Note that only the 
field $\hat{a}_{in}$ appears on the right side of the equation, because we assume 
without loss of generality that we inject a tone on the signal side 
(not on the idler side) of the input field. 
Similarly one could describe an input tone 
injected on the idler side by considering the adjoint of the same equation.
This we will describe in more detail further below. Recalling again 
the dimension of $\hat{a}_{in}$ of $\sqrt{s}$, the right side side of Eq.~(\ref{eq:S21}) 
has the dimension $1/s$, like the left side.  
Exclusively for the Fourier 
transformation of Eq.~(\ref{eq:S19}) we use the convention 
$f(t) = \int d\omega F(\omega) exp(-i\omega t)$ and 
$F(\omega) = 1/(2\pi)\int dt f(t) exp(+i\omega t)$
bringing the expression (\ref{eq:S21}) to a tractable form.

From now on we will identify $\omega = \omega_{s}$, where $\omega_{s}$ is the signal
frequency and we will use the symbol $\omega_{i}$ to denote the idler frequency of the 
parametric amplifier. Equally we will use the indices '$s$' and '$i$' to label some 
operators we will use in the following which act on the signal and idler frequencies.

Our goal is now to bring Eq.~(\ref{eq:S21}) into a form which is compatible 
with the input-output formalism.
In particular Eq.~(\ref{eq:S21}) should assume a form which allows us 
to compare it with the input-output relation for a simple harmonic 
oscillator (without nonlinearity from a Josephson junction or other 
nonlinear elements), which we would like to use for a sanity check and to 
highlight the effect of the nonlinearity in our system. 

We proceed to rewrite Eq.~(\ref{eq:S21}) by performing a 
rotating-wave-approximation so that the equation rotates approximately at the pump frequency. 
For this we also apply the following substitutions, $\omega_{p} + \omega_{s} \approx 2\omega_{p}$
and also $\omega_{s}/\omega_{0} \approx 1$. This means essentially also that 
around the resonator's plasma frequency we can assume that 
$\omega_{s}\approx \omega_{p} \approx \omega_{0}$ to good approximation.

We first divide Eq.~(\ref{eq:S21}) by $\omega_{0}$ and then add and subtract the adequate amount of 
frequencies so that at the end we can divide by the factor of '2' in front of the term proportional to 
$\kappa[\omega]$ on the left side of Eq~(\ref{eq:S21}) without introducing artificially fractions of frequencies. 
We obtain then:
\begin{widetext}
\begin{equation}
\begin{split}
\label{eq:S23}
&\left[-\omega_{s}^{2} - i\omega_{s}\kappa[\omega_{s}] 
+ \omega_{0}^{2}\left(1-\frac{\delta_{p,max}^{2}}{4}\right)\right]\hat{\delta}_{s}[\omega_{s}]
e^{-i\varphi}
- \frac{\omega_{0}^{2}\delta_{p,max}^{2}}{8}
\hat{\delta}^{\dagger}_{s}\left[-\omega_{s} + 2\omega_{p}\right]e^{+i\varphi}
=\frac{4\pi \kappa[\omega_{s}]}{\phi_{0}}\sqrt{\frac{\hbar \omega_{s}}{2Y''''_{in}[\omega_{s}]}}
\hat{a}_{in}[\omega_{s}]e^{-i\varphi}\\
\Rightarrow&
\left[\omega_{0} - \omega_{s} - \frac{\omega_{0}\delta_{p,max}^{2}}{4} - i\kappa[\omega_{s}]\right]
\hat{\delta}_{s}[\omega_{s}]e^{-i\varphi}
- \frac{\omega_{0}\delta_{p,max}^{2}}{8}
\hat{\delta}^{\dagger}_{s}\left[-\omega_{s} + 2\omega_{p}\right]e^{+i\varphi}
=\frac{4\pi \kappa[\omega_{s}]}{\phi_{0}\omega_{0}}\sqrt{\frac{\hbar \omega_{s}}{2Y''''_{in}[\omega_{s}]}}
\hat{a}_{in}[\omega_{s}]e^{-i\varphi}\\
\Rightarrow&
\left[2\tilde{\omega}_{p} -
2\tilde{\omega}_{s} - \frac{\omega_{0}\delta_{p,max}^{2}}{4} - i\kappa[\tilde{\omega}_{s}]\right]
\hat{\delta}_{s}[\tilde{\omega}_{s}]e^{-i\varphi}
- \frac{\omega_{0}\delta_{p,max}^{2}}{8}
\hat{\delta}^{\dagger}_{s}\left[-\tilde{\omega}_{s}\right]e^{+i\varphi}
=\frac{4\pi \kappa[\tilde{\omega}_{s}]}{\phi_{0}\omega_{0}}\sqrt{\frac{\hbar \omega_{s}}{2Y''''_{in}[\tilde{\omega}_{s}]}}
\hat{a}_{in}[\tilde{\omega}_{s}]e^{-i\varphi}\\
\Rightarrow&
\left[\tilde{\omega}_{p} -
\tilde{\omega}_{s} - \frac{\omega_{0}\delta_{p,max}^{2}}{8} - 
\frac{i\kappa[\tilde{\omega}_{s}]}{2}\right]
\hat{\delta}_{s}[\tilde{\omega}_{s}]e^{-i\varphi}
- \frac{\omega_{0}\delta_{p,max}^{2}}{16}
e^{+i\varphi}\hat{\delta}^{\dagger}_{s}\left[-\tilde{\omega}_{s}\right]
=\frac{2\pi \kappa[\tilde{\omega}_{s}]}{\phi_{0}\omega_{0}}\sqrt{\frac{\hbar \omega_{s}}{2Y''''_{in}[\tilde{\omega}_{s}]}}
\hat{a}_{in}[\tilde{\omega}_{s}]e^{-i\varphi}\\
\Rightarrow&
\left[i(\tilde{\Omega}_{p} -
\tilde{\omega}_{s}) + \frac{\kappa[\tilde{\omega}_{s}]}{2}\right]
\hat{\delta}_{s}[\tilde{\omega}_{s}]e^{-i\varphi}
- \frac{i\omega_{0}\delta_{p,max}^{2}}{16}
e^{+i\varphi}\hat{\delta}^{\dagger}_{s}\left[-\tilde{\omega}_{s}\right]
=i\frac{2\pi \kappa[\tilde{\omega}_{s}]}{\phi_{0}\omega_{0}}\sqrt{\frac{\hbar \omega_{s}}{2Y''''_{in}[\tilde{\omega}_{s}]}}
\hat{a}_{in}[\tilde{\omega}_{s}]e^{-i\varphi}~.
\end{split}
\end{equation}
\end{widetext}
We use in the derivation above the notation for the pump and signal detuning, indicated by a \emph{tilde} 
like in the main text and as mentioned in the beginning 
of this supplemental information; $\tilde{\omega}_{p} = \omega_{0} - \omega_{p}$ and 
$\tilde{\omega}_{s} = \omega_{p} - \omega_{s}$. We denote the effective parametric pump frequency 
with a capital $\tilde{\Omega}_{p}$ since it contains the pump frequency detuning 
$\tilde{\omega}_{p}$ and the additional term $-\omega_{0}\delta_{p,max}^{2}/8$, leading 
to the frequency shift we have already shown in Fig.~\ref{figS01}; 
$\tilde{\Omega}_{p} = \tilde{\omega}_{p} - \omega_{0}\delta_{p,max}^{2}/8$.
This is a typical characteristic of a nondegenerate parametric amplifier in 
which one pumps a Kerr-type nonlinearity through the signal port; 
the oscillator response shifts when the pumping power increases. 
This has to be taken into account during the operation of the 
device and an amplifier design is desired where this effect 
is minimal.

Note also that the negative frequency argument
in the expression $\hat{\delta}^{\dagger}_{s}\left[-\tilde{\omega}_{s}\right]$ means that 
a photon on the idler side is created with frequency $\omega_{i} = 2\omega_{p} - \omega_{s}$ and 
$\tilde{\omega}_{s} \geq 0$.

We now go over to creation and annihilation fields by applying the following identities to 
Eq.~(\ref{eq:S23}), compatible with the input-output formalism as 
described by [A.~A.~Clerk et al., Rev.~Mod.~Phys.~{\bf 82}, 1155, (2010)]; 
$\hat{a}_{s}[\tilde{\omega}_{s}] = \hat{\delta}_{s}[\tilde{\omega}_{s}]e^{-i\varphi}$, 
$\hat{a}_{i}[-\tilde{\omega}_{s}] = \hat{\delta}_{s}^{\dagger}[-\tilde{\omega}_{s}]e^{+i\varphi}$ and 
$i\frac{2\pi \kappa[\tilde{\omega}_{s}]}{\phi_{0}\omega_{0}}
\sqrt{\frac{\hbar \omega_{s}}{2Y''''_{in}[\tilde{\omega}_{s}]}}
\hat{a}_{in}[\tilde{\omega}_{s}]e^{-i\varphi} \rightarrow \hat{a}_{in}[\tilde{\omega}_{s}]$, 
where now the indices $s$ and $i$ of the intra-oscillator 
fields $\hat{a}, \hat{a}^{\dagger}$ label
specifically the signal (s) and idler (i) modes. Note that after these variable transformations,  
$\hat{a}_{in}$ is dimensionless, whereas the intra-oscillator fields obtain a dimension of $s$.

Equation~(\ref{eq:S23}) then assumes the simple form: 
\begin{equation}
\label{eq:S24}
\left[i(\tilde{\Omega}_{p} -
\tilde{\omega}_{s}) + \frac{\kappa[\tilde{\omega}_{s}]}{2}\right]
\hat{a}_{s}[\tilde{\omega}_{s}]
- \frac{i\omega_{0}\delta_{p,max}^{2}}{16}
\hat{a}_{i}^{\dagger}[-\tilde{\omega}_{s}]
= \hat{a}_{in}[\tilde{\omega}_{s}]~.
\end{equation}
Together with its adjoint (effectively this interchanges 
the role of signal and idler terms):
\begin{equation}
\begin{split}
\label{eq:S25}
\left[-i(\tilde{\Omega}_{p} +
\tilde{\omega}_{s}) + \frac{\kappa^{*}[-\tilde{\omega}_{s}]}{2}\right]
\hat{a}_{i}^{\dagger}[-\tilde{\omega}_{s}]
+ \frac{i\omega_{0}\delta_{p,max}^{2}}{16}
&\hat{a}_{s}[\tilde{\omega}_{s}]\\
&= \hat{a}_{in}^{\dagger}[-\tilde{\omega}_{s}]~,
\end{split}
\end{equation}
where again $\tilde{\omega}_{s} \geq 0$. We obtain all coefficients 
necessary to build up the susceptibility matrix for the 
parametric amplifier. This matrix is 
needed for the description of the parametric amplifier in the framework 
of the input-output formalism which is briefly sketched in the following 
section and which we have heavily condensed for the purposes in this paper.
\section*{Input-output formalism and quantum noise for a nondegenerate 
(phase preserving) JPA}
With the previous preparation at hand we formulate a relation which connects the 
intra-oscillator field $\hat{\mathbf{a}} = \left(\hat{a}_{s}[\tilde{\omega}_{s}], 
\hat{a}_{i}^{\dagger}[-\tilde{\omega}_{s}]\right)^{T}$ with the 
input field $\hat{\mathbf{a}}_{in} = \left(\hat{a}_{in}[\tilde{\omega}_{s}], 
\hat{a}_{in}^{\dagger}[-\tilde{\omega}_{s}]\right)^{T}$ 
in consistence with the work done by [A.~A.~Clerk et al., 
Rev.~Mod.~Phys.~{\bf 82}, 1155, (2010);  
C.~Laflamme and A.~A.~Clerk, Phys.~Rev.~A~{\bf 83}, 033803, (2011); 
Tanay Roy et al., Appl.~Phys.~Lett.~{\bf 107}, 262601, (2015); 
R.~Vijay et al., Rev.~Sci.~Instrum.~{\bf 80}, 111101, (2009) and 
V.~E.~Manucharyan et al.~Phys.~Rev.~B~{\bf 76}, 014524, (2007)]:
\begin{equation}
\label{eq:S26}
\chi[\tilde{\omega}_{s}]^{-1}\cdot\hat{\bf{a}}[\tilde{\omega}_{s}] = 
\hat{\bf{a}}_{in}[\tilde{\omega}_{s}]~.
\end{equation}
The inverted susceptibility matrix [A.~A.~Clerk et al., 
Rev.~Mod.~Phys.~{\bf 82}, 1155, (2010); 
C.~Laflamme and A.~A.~Clerk, Phys.~Rev.~A~{\bf 83}, 
033803, (2011)] reads:
\begin{equation}
\label{eq:S27}
\chi[\tilde{\omega}_{s}]^{-1}
=
-i\begin{psmallmatrix}
\left[-(\tilde{\Omega}_{p} -
\tilde{\omega}_{s}) + \frac{i\kappa[\tilde{\omega}_{s}]}{2}\right]
&\frac{\omega_{0}\delta_{p,max}^{2}}{16}\\
- \frac{\omega_{0}\delta_{p,max}^{2}}{16}
&
\left[(\tilde{\Omega}_{p} +
\tilde{\omega}_{s}) + \frac{i\kappa^{*}[-\tilde{\omega}_{s}]}{2}\right]
\end{psmallmatrix}~,
\end{equation}
being the coefficient matrix of the equation system (\ref{eq:S24}) 
and (\ref{eq:S25}). As mentioned in the previous section, we now 
want to compare the susceptibility (\ref{eq:S27}) with the one
for an one-sided empty oscillator without nonlinearity. In this case we find a similar 
susceptibility just without off-diagonal entries which are on the other hand 
indispensable in parametric amplification in order to enable 
energy exchange between the frequency modes.

Following the theory in [A.~A.~Clerk et al., 
Rev.~Mod.~Phys.~{\bf 82}, 1155, (2010); 
C.~Laflamme and A.~A.~Clerk, Phys.~Rev.~A~{\bf 83}, 
033803, (2011)] the photon number 
gain (or power gain) of the signal mode of the 
parametric amplifier can then be evaluated by evaluating the 
following relation:
\begin{equation}
\label{eq:S28}
G_{s}[\tilde{\omega}_s] = 
\lvert1 - \kappa[\tilde{\omega}_{s}]\chi_{11}[\tilde{\omega}_{s}]\rvert^{2}~,
\end{equation}
and the photon number gain of the idler mode reads:
\begin{equation}
\label{eq:S29}
G_{i}[-\tilde{\omega}_s] = 
\lvert1 - \kappa[-\tilde{\omega}_{s}]\chi_{11}[-\tilde{\omega}_{s}]\rvert^{2}~.
\end{equation}
For the evaluation, we numerically invert (\ref{eq:S27}) for each 
detuning $\tilde{\omega}_{s}$ and subsequently substitute the 
matrix element $\chi_{11}$ into the equations (\ref{eq:S28}) and (\ref{eq:S29}). 
For a symmetric admittance function around $\tilde{\omega}_{s}$ like in our work 
we obviously find $G_{s} = G_{i}$. Asymmetries are introduced by other imperfections 
of the circuit which we detail in the main text.

While the susceptibility matrix (\ref{eq:S27}) connects the intra-oscillator with 
the input field, yielding the gain of the parametric amplifier, 
supplemental input-output relations connect further the 
input fields with the output (amplified) fields:
\begin{equation}
\label{eq:S30}
\begin{pmatrix}
\hat{a}_{out}[\tilde{\omega}_{s}]\\ \hat{a}_{out}^{\dagger}[-\tilde{\omega}_{s}]
\end{pmatrix}
=
\begin{pmatrix}
\mathcal{G}[\tilde{\omega}_{s}]&\mathcal{M}[\tilde{\omega}_{s}]
\\
\mathcal{M}^{*}[-\tilde{\omega}_{s}]&\mathcal{G}^{*}[-\tilde{\omega}_{s}]
\end{pmatrix}
\begin{pmatrix}
\hat{a}_{in}[\tilde{\omega}_{s}]\\ \hat{a}_{in}^{\dagger}[-\tilde{\omega}_{s}]
\end{pmatrix}~.
\end{equation}
This gain relation of an idealized and initially 
lossless nondegenerate parametric amplifier 
is of particular importance to understand the minimum noise which is 
added by a real amplifier and will 
lead to the fundamental result of the 
Haus-Caves theorem [H.~A.~Haus and J.~A.~Mullen, 
Phys.~Rev.~{\bf 128}, 2407, (1962); 
C.~C.~Caves, Phys.~Rev.~D~{\bf 26}, 1817, (1982)].
 
Here, the calligraphic letters in the gain matrix 
denote the amplitude gain instead of the power gain which is on the other hand 
evaluated by Eqs.~(\ref{eq:S28}) and (\ref{eq:S29}). Specifically, 
$\mathcal{G}[\tilde{\omega}_{s}]$ denotes the amplitude 
gain at the signal frequency whereas $\mathcal{M}[-\tilde{\omega}_{s}]$ denotes 
the amplitude conversion gain at the idler frequency for a frequency being injected 
into the parametric amplifier at the signal frequency $\tilde{\omega}_{s}$.

The following identities are needed for the further presentation:
\begin{equation}
\begin{split}
\label{eq:S31}
&\lvert \mathcal{G}[\pm \tilde{\omega}_{s}] \rvert^{2} - 
\lvert \mathcal{M}[\pm \tilde{\omega}_{s}] \rvert^{2} = 1\\
&\mathrm{and}\\
&\mathcal{G}[\tilde{\omega}_{s}]\mathcal{M}[-\tilde{\omega}_{s}] = 
\mathcal{G}[-\tilde{\omega}_{s}]\mathcal{M}[\tilde{\omega}_{s}]~.
\end{split}
\end{equation}
The power gain can then be expressed via the amplitude 
gain as $G_{s}[\tilde{\omega}_{s}] = 
\lvert \mathcal{G}[\tilde{\omega}_{s}]\rvert^{2}$ and 
$G_{i}[-\tilde{\omega}_{s}] = 
\lvert \mathcal{M}[\tilde{\omega}_{s}]\rvert^{2}$ and 
$G_{i}[-\tilde{\omega}_{s}] = G_{s}[\tilde{\omega}_{s}] - 1$. Clearly, for large power 
gains $\gg 1$, $G_{s} \approx G_{i}$. 

We now go over to an equation like Eq.~(\ref{eq:S30}) including 
noise and want to focus 
on the ground state of the electromagnetic field. By this we 
determine the minimum noise the nondegenerate
parametric amplifier will add to the signal referred to the input. 
For this we will consider a blackbody 
load at temperature $k_{B} T \ll \hbar \omega_{p}$, i.e.~close to its ground state, 
injecting power into the parametric amplifier and we calculate the power 
spectral density at its output.

The first component of Eq.~(\ref{eq:S30}) reads 
$\hat{a}_{out}[\tilde{\omega}_{s}] 
= \mathcal{G}[\tilde{\omega}_{s}] \hat{a}_{in}[\tilde{\omega}_{s}] + 
\mathcal{M}[\tilde{\omega}_{s}]\hat{a}^{\dagger}_{in}[-\tilde{\omega}_{s}]$ 
and shows two important properties of 
a nondegenerate JPA. First, signal and idler modes are strongly 
correlated since $\mathcal{G}\approx\mathcal{M}$ for high gains 
and, second, noise from the idler is added to the signal portion of the amplified signal.
We will be more specific on the second property now.

In order to simplify the following discussion, we use the fact that a nondegenerate 
JPA does not favor the amplification/de-amplification of a 
particular signal quadrature, rather it treats both quadratures equally and 
independent of the input signal phase shift with respect to the pump tone.

The total noise of the JPA at its output with respect 
to its input (compare with Fig.~\ref{figS02}) can be expressed by quantifying the
fluctuations of the input field and how they are transformed into the output field.
We write the first component of Eq.~(\ref{eq:S30}) in terms of noise power spectral 
densities of the input signal and idler and of the output fields and obtain:
\begin{widetext}
\begin{equation}
\begin{split}
\label{eq:S32}
\mathcal{S}_{out}[\tilde{\omega}_{s}] &= 
G_{s}[\tilde{\omega}_{s}]\mathcal{S}_{in}[\tilde{\omega}_{s}] 
+ G_{i}[-\tilde{\omega}_{s}]\mathcal{S}_{in}[-\tilde{\omega}_{s}]\\
&=G_{s}[\tilde{\omega}_{s}]\mathcal{S}_{in}[\tilde{\omega}_{s}] 
+ \left(G_{s}[\tilde{\omega}_{s}] - 1\right)\mathcal{S}_{in}[-\tilde{\omega}_{s}]\\
&=G_{s}[\tilde{\omega}_{s}]\frac{\hbar\omega_{p}}{2}\left( 
\langle\hat{a}_{in}[\tilde{\omega}_{s}]\hat{a}_{in}^{\dagger}[\tilde{\omega}_{s}]\rangle 
+ \langle\hat{a}_{in}^{\dagger}[\tilde{\omega}_{s}]\hat{a}_{in}[\tilde{\omega}_{s}]\rangle\right)
+ \left(G_{s}[\tilde{\omega}_{s}] - 1\right) \frac{\hbar\omega_{p}}{2}
\left(\langle\hat{a}_{in}[-\tilde{\omega}_{s}]\hat{a}_{in}^{\dagger}[-\tilde{\omega}_{s}]\rangle 
+ \langle\hat{a}_{in}^{\dagger}[-\tilde{\omega}_{s}]\hat{a}_{in}[-\tilde{\omega}_{s}]\rangle\right)\\
&=G_{s}[\tilde{\omega}_{s}] \frac{\hbar\omega_{p}}{2}
\coth\left(\frac{\hbar\omega_{p}}{2k_{B}T}\right) + 
\left(G_{s}[\tilde{\omega}_{s}] - 1\right) \frac{\hbar\omega_{p}}{2}
\coth\left(\frac{\hbar\omega_{p}}{2k_{B}T}\right)\\
&=G_{s}[\tilde{\omega}_{p}]\Bigg\lbrace\frac{\hbar\omega_{p}}{2}
\coth\left(\frac{\hbar\omega_{p}}{2k_{B}T}\right) + 
\left(1-\frac{1}{G_{s}[\tilde{\omega}_{s}]}\right)
\frac{\hbar\omega_{p}}{2}\coth\left(\frac{\hbar\omega_{p}}{2k_{B}T}\right)\Bigg\rbrace~.
\end{split}
\end{equation}
\end{widetext}
where as said before, the $G$'s are the power gains. Also we assume that the signal 
and idler frequencies are close enough to the pump frequency $\omega_{p}$ so that 
we can develop the noise spectral densities around $\omega_{p}$ without introducing 
too large errors. By considering the small 
change of the $\coth(\cdot)$-terms in the $\mathrm{-3~dB}$ bandwidth 
of our amplifier this approximation is obviously valid for our case. The last line of 
Eq.~(\ref{eq:S32}) is the fundamental result of the Haus-Caves theorem for a 
nondegenerate JPA (or in other words a phase-preserving amplifier), which for 
gains $\gg 1$ will add a minimum amount of shot noise equivalent to half a photon 
per second per Hertz of bandwidth to the amplified 
portion of the signal at the input. In total, together with the vacuum fluctuation in the signal 
input field, the nondegenerate JPA will amplify at least a shot noise equivalent 
to one photon per second per Hertz of bandwidth 
at its input which is added to the desired signal in the input field.

For a real device, the noise as a function of the signal frequency is not flat 
which on the other hand is suggested by 
Eq.~(\ref{eq:S32}). The gain profile, the coupling to the device, the shape of the 
input admittance function $Y_{in}''''[\tilde{\omega}_{s}]$ and possible losses 
in the signal path will contribute to the total noise with respect to the input and will 
increase the minimum noise amount suggested by Eq.~(\ref{eq:S32}). 
We describe this in detail in the main text for our device.
\section*{Broadband dielectric loaded coplanar waveguide-to-microstrip 
transformer for quantum circuits}
\begin{figure}[tb]
\centering
\includegraphics[width=\columnwidth]{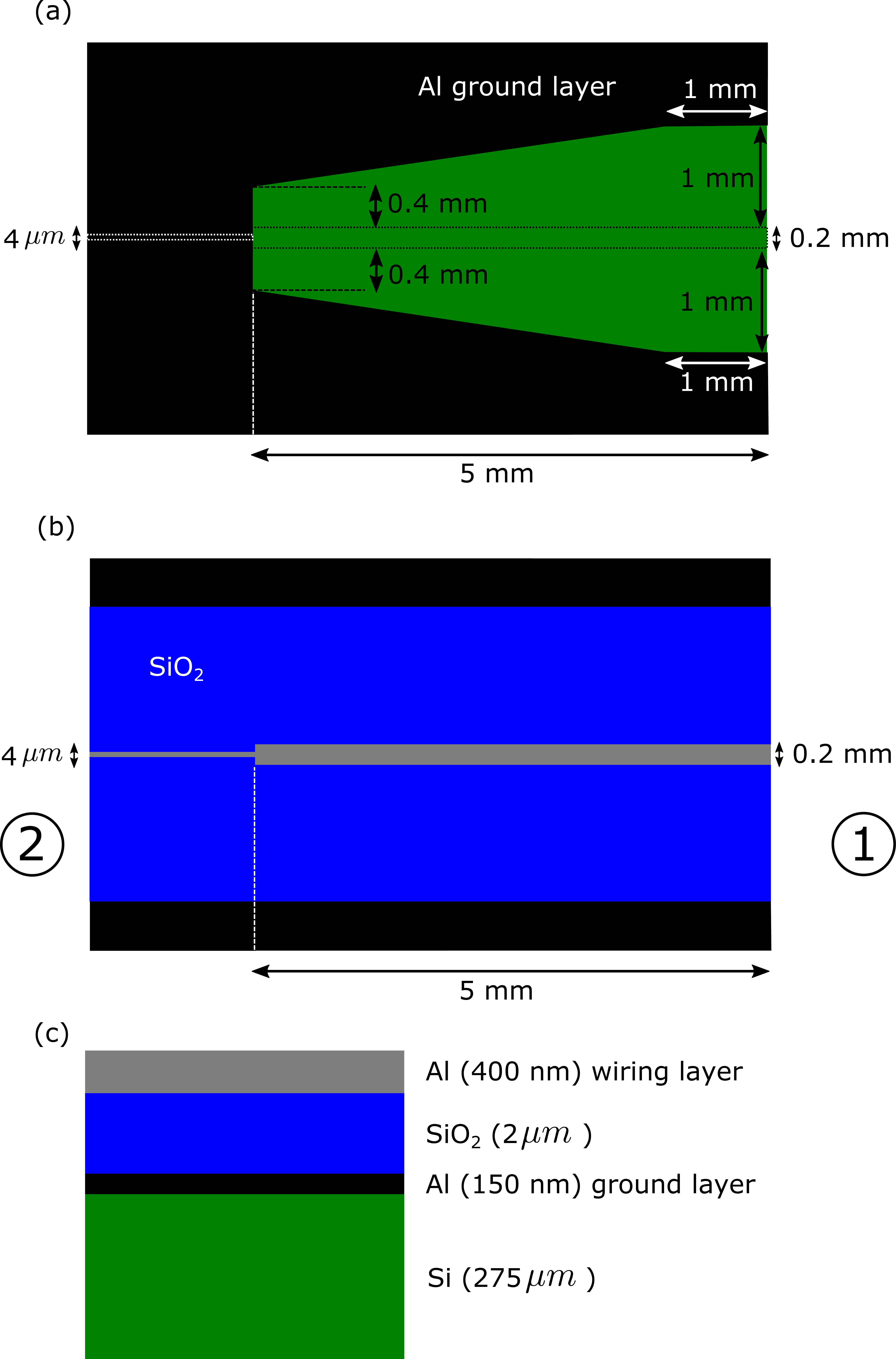}
\caption{\label{figS03}(a) and (b), dimensions 
for an ultra broadband dielectric loaded CPW-to-MS 
transformer where port '2' is the one which could connect to the JPA 
microstrip circuitry. Both ports '1' and '2' have a characteristic impedance of $50~\Omega$. 
(a) shows the ground layer and (b) shows the wiring layer 
on top of a $\mathrm{SiO_{2}}$ dielectric layer for which we assume 
a dielectric constant of $\epsilon \approx 3.75$. In (a) we indicate 
by the dotted lines the dimension and position of the wiring layer shown in (b). 
(c) Layer stack of the circuit. Note that while the bulk Si handler wafer 
is also the dielectric for the dielectric loaded CPW, 
the relevant dielectric layer for the MS circuit is 
just the $\mathrm{SiO_{2}}$ layer and for this part of the circuit 
the fields are confined between the ground (shown in black color) 
and wiring layer (shown in grey color).}
\end{figure}

\begin{figure}[tb]
\centering
\includegraphics[width=\columnwidth]{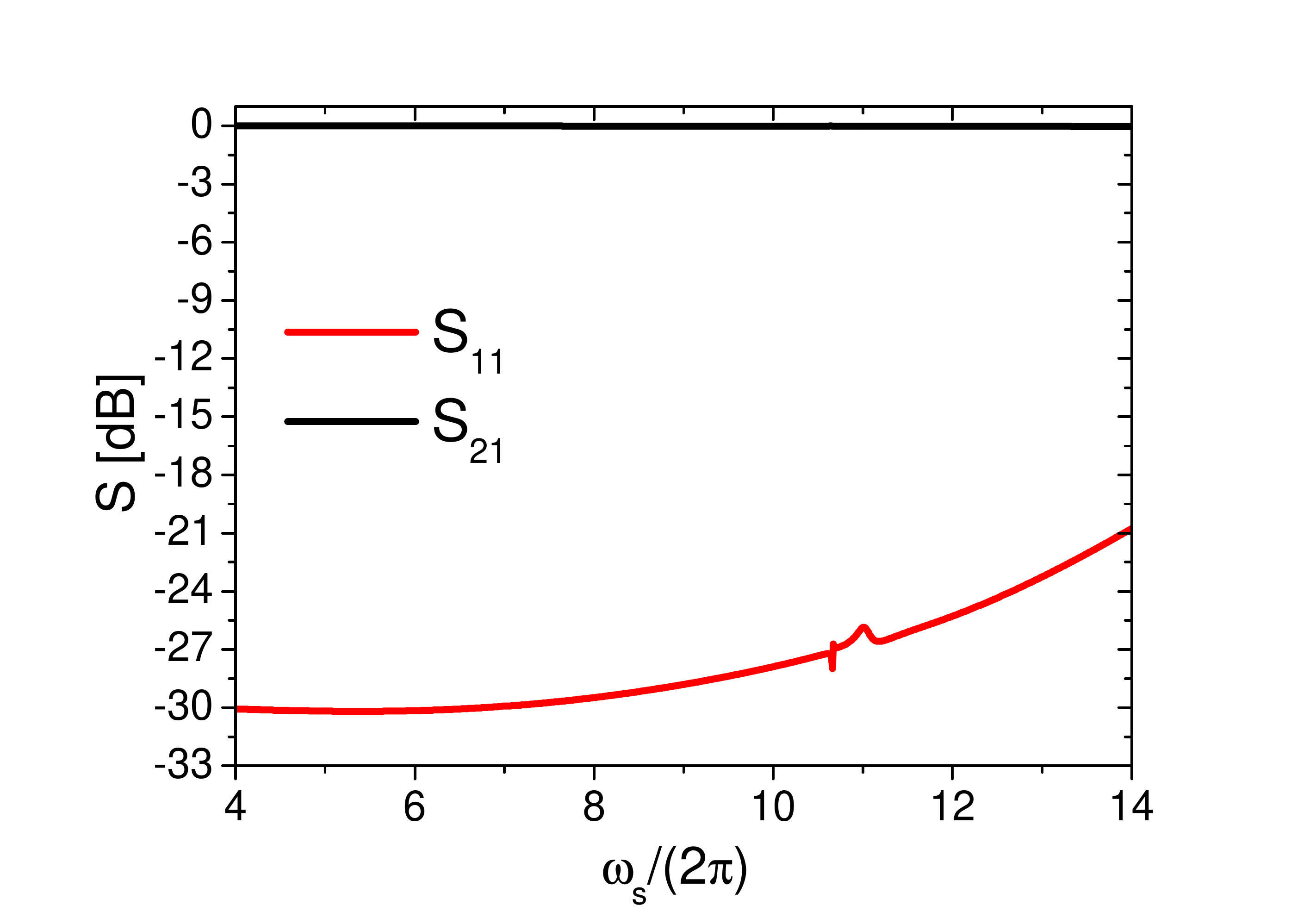}
\caption{\label{figS04}Simulation results for the 
circuit shown in Fig.~\ref{figS03}. We show the scattering 
parameters for excitation from port '1' whereas the 
simulation obtains the same 
scattering parameters for excitation from port '2' 
because of the reciprocity of the circuit.}
\end{figure}

In this section we provide design dimensions and simulation results obtained with 
CST microwave studio [CST-Computer Simulation Technology, \emph{https://www.cst.com}] 
for an ultra broadband dielectric loaded coplanar waveguide 
(CPW)-to-microstrip (MS) transformer circuit. 
With such a transformer, one could connect our microstrip amplifier to a CPW structure, either 
for connection to the input/output cabling or to other quantum circuits which are often 
designed using CPW geometries. Figures~\ref{figS03} and \ref{figS04} 
present our results. Both ports '1' and '2' have a characteristic impedance of $50~\Omega$.
\section*{Dispersion relation on a superconducting Al microstrip transmission line}
We employ aluminum based superconducting microstrip (MS) transmission lines 
in our device. It is known that the geometrical length corresponding to some 
electrical length on such a transmission line (and in principle 
no matter which material is employed) is shorter than in its normal conducting state
[D.~C.~Mattis and J.~Bardeen, Phys.~Rev.~{\bf 111}, 412, (1958); 
R.~L.~Kautz, J.~Appl.~Phys.~{\bf 49}, 308, (1978)]. The fundamental reason for this is 
that the phase velocity slows down when the material becomes superconducting 
[R.~L.~Kautz, J.~Appl.~Phys.~{\bf 49}, 308, (1978)] 
and to compensate for that one has to shorten the transmission line.
There are three leading influences how strong the dispersion relation, and therefore the 
phase velocity, is modified in the superconducting state of the MS transmission 
line compared to its normal-state value.

First, the geometry of the MS transmission line, 
in particular the ratio between MS conductor
width $W$ and dielectric thickness $d$, second, the normal state resistivity $\rho_{n}$ 
(measured for Al at 77~K) and finally the superconducting gap $\Delta$. 
For large $\rho_{n}$ and small $\Delta$, the shortening of the geometrical 
length which corresponds to some fixed electrical length on the superconducting 
microstrip transmission line
compared to the normal-conducting state is most pronounced. This 
shortening becomes less severe when $\rho_{n}$ becomes 
smaller and smaller like it is the case in high-quality Al.

In our work we propose microstrip transmission line geometries yielding 
characteristic impedances of the order of $50~\Omega$ or less 
using a $2~\mu\mathrm{m}$ thick dielectric layer made of 
$\mathrm{SiO_{2}}$, for which we assume a relative 
permittivity of $\epsilon_{r} \approx 3.75$. This makes a microstrip 
transmission line geometry necessary where $W/d > 1$, but 
which is not much larger than '1', like required for the 
approximations in the work of 
Kautz [R.~L.~Kautz, J.~Appl.~Phys.~{\bf 49}, 308, (1978)]. 
Therefore, fringing fields play an increasing role in our case and a 
conformal mapping calculation keeping higher order terms to 
obtain more precise values for the capacitive admittance and series 
impedance per unit length line of the 
transmission line becomes necessary.

\emph{Collin} reports such a conformal 
mapping technique using a similar microstrip geometry liked employed in our design 
[R.~E.~Collin, 'Foundations for Microwave Engineering', 2nd edition, IEEE Press, Wiley-Interscience]. The theory provided by \emph{Kautz} can then be 
modified to account for the microstrip transmission line geometries in our device. 

The capacitive admittance and series impedance including higher order terms as 
a result of the 
conformal mapping calculation reads then:
\begin{equation}
\begin{split}
\label{eq:S33}
Y[\omega] &= i\omega\epsilon_{0}\epsilon_{eff}
\left[\frac{W}{d} + 1.393 + 0.667\ln\left(\frac{W}{d} + 1.444\right)\right]\\
Z[\omega] &= i\omega \mu_{0} 
\left[\frac{W}{d} + 1.393 + 0.667\ln\left(\frac{W}{d} + 1.444\right)\right]^{-1}\\ 
&+ \frac{1}{W}\left(Z_{s,b}[\omega] + Z_{s,w}[\omega]\right)~.\\
\end{split}
\end{equation}
In the above equations $\epsilon_{0}$ and $\mu_{0}$ are the free space permittivity and 
permeability and the effective dielectric constant reads:
\begin{equation}
\begin{split}
\label{eq:S34}
\epsilon_{eff} &= \frac{\epsilon_{r} + 1}{2} + \frac{\epsilon_{r} - 1}{2}
\left(1+\frac{12d}{W}\right)^{-1/2} \\
&- 0.217\left(\epsilon_{r}-1\right)\frac{t}{\sqrt{Wd}}~,
\end{split}
\end{equation}
where $t$ is the thickness of the wiring conductor of the microstrip transmission line.
Equations~(\ref{eq:S33}) and (\ref{eq:S34}) are valid for the case $W>d$. 
Furthermore, $Z_{s,g}$ and $Z_{s,w}$
are the surface impedances of the ground and wiring layers of the superconducting 
Al microstrip transmission line. For the latter two quantities and 
for all practically relevant situations in Al one can assume 
the local and dirty limit (the magnetic penetration depth is larger or of the order of the 
coherence length and the coherence length is larger than the electron mean free path), 
although one has to realize that Al is pretty much at the 
transition between the local and dirty limit and the extreme anomalous limit 
(the magnetic penetration depth is smaller than the coherence length and 
than the electron mean free path) 
[D.~C.~Mattis and J.~Bardeen, Phys.~Rev.~{\bf 111}, 412, (1958)]. 
However, in practice the local and dirty limits account well enough for the experimental 
observations and greatly simplify the dispersion relation calculations. 
The expression for the surface impedance in these limits then reads:
\begin{equation}
\begin{split}
\label{eq:S35}
Z[\omega] = \sqrt{\frac{i\omega\mu_{0}}{\sigma[\omega]}}
\coth{\left(\sqrt{i\omega\mu_{0}\sigma[\omega]}t\right)}~,
\end{split}
\end{equation}
where $\sigma[\omega] = \sigma_{1}[\omega] - i\sigma_{2}[\omega]$ is the complex conductivity 
from the Mattis-Bardeen theory 
[D.~C.~Mattis and J.~Bardeen, Phys.~Rev.~{\bf 111}, 412, (1958)] 
and $t$ is the conductor thickness
of the ground and wiring layers as indicated in Fig.~\ref{fig01}(c) 
of the main text and in Fig.~\ref{figS03}(c) of this supplemental 
information. Equation~(\ref{eq:S35}) depends on the superconducting gap through 
the complex conductivity 
[D.~C.~Mattis and J.~Bardeen, Phys.~Rev., {\bf 111}, 412, (1958)].

The characteristic impedance can then be expressed as usual as:
\begin{equation}
\label{eq:S36}
Z_{c}[\omega] = \sqrt{\frac{Z[\omega]}{Y[\omega]}}~,
\end{equation}
and the complex propagation constant 
$\gamma[\omega] = \alpha[\omega] + i\beta[\omega]$ reads:
\begin{equation}
\label{eq:S37}
\gamma[\omega] = \sqrt{Z[\omega]Y[\omega]}~.
\end{equation}
The loss and geometrical length for one wavelength 
on the transmission line is evaluated as:
\begin{equation}
\begin{split}
\label{eq:S38}
\alpha[\omega] &= \mathrm{Re}\left(\gamma[\omega]\right)\\
\lambda[\omega] &= \frac{2\pi}{\mathrm{Im}\left(\gamma[\omega]\right)}~.
\end{split}
\end{equation}

In a final step we compare circuit simulations using CST microwave studio 
[CST-Computer Simulation Technology, \emph{https://www.cst.com}] from which 
we obtain the dispersion relation on the superconducting Al 
microstrip transmission line with the 
dispersion relation we obtain from the analytical theory we have presented before. 
In the model we establish in CST, we assume an ultra low loss electrical 
(normal) conductor for reasons of computational speed and model 
simplicity. We find that 
for a low normal-state resistivity of the superconducting 
Al material of $0.1~\mu\Omega\mathrm{cm}$, 
the CST model reproduces the results of the analytical theory 
(taking superconductivity explicitly
into account) up to an uncertainty of 3\% for both 
the characteristic impedance and for the geometrical 
length corresponding to an electrical length of one wavelength 
on the superconducting microstrip transmission line. For higher resistivities of say 
$0.3~\mu\Omega\mathrm{cm}$ and $0.6~\mu\Omega\mathrm{cm}$, the 
characteristic impedance which we obtain from the same CST model is still 
in reasonable agreement with the analytical theory. However, the 
analytical theory yields a geometrical length, corresponding to one 
wavelength on the superconducting transmission line, which is 
shorter by 4\% and by 6\% 
compared to the normal conducting state (or compared to a superconducting 
Al conductor with very low normal-state resistivity). 
In all of these calculations we assume a superconducting gap 
of $\Delta = 0.182$~meV at 10~mK.
 
We have observed similar, more severe effects 
in our earlier works [M.~P.~Westig et al., 
J.~Appl.~Phys.~{\bf 114}, 124504, (2013); 
M.~P.~Westig et al., J.~Appl.~Phys.~{\bf 112}, 093919, (2012); 
M.~P.~Westig et al., Supercond.~Sci.~Technol.~{\bf 24}, 085012, (2011)] 
using niobium based (and, hence, a material 
with up to ten times higher normal-state resistivity than in Al)
microstrip circuits. Circuits which should employ the disordered superconductors 
NbN or NbTiN will lead to an even more severe modification of the dispersion relation, 
resulting in an effective geometric shortening of the transmission line 
(again with the goal to keep the electrical length constant) of up to 30\% because of 
normal-state resistivities approaching $100~\mu\Omega\mathrm{cm}$ and more.
\section*{Additional circuit characteristics}
\begin{figure}[htb]
\centering
\includegraphics[width=\columnwidth]{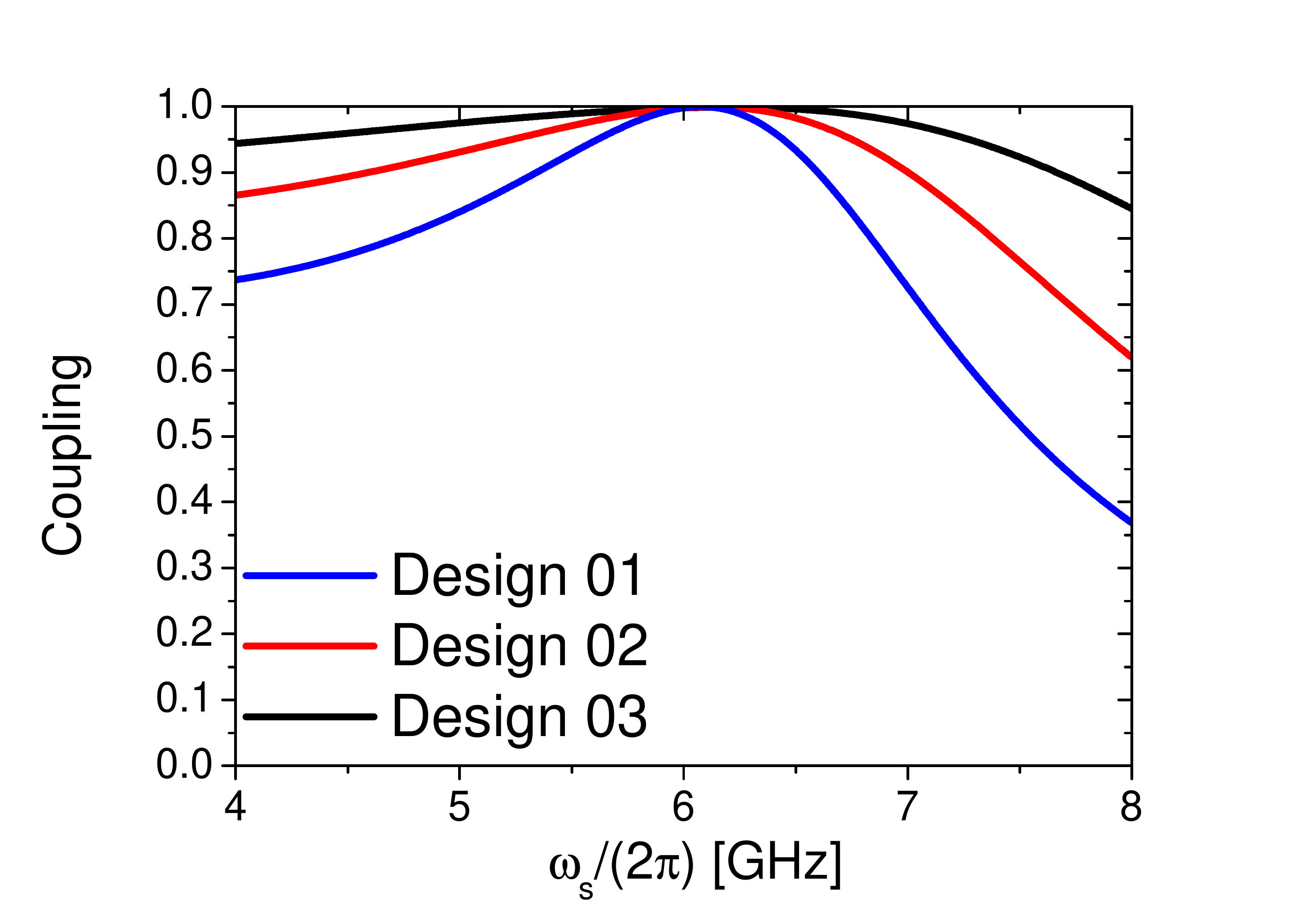}
\caption{\label{figS05}Coupling from the embedding circuit to the 
Josephson junction oscillators referred to the point where we specify $Y_{in}''''$ 
in Fig.~\ref{fig01}(a). For all three designs, the 
coupling is between '1' and '0.95'.}
\end{figure}

\begin{figure}[htb]
\centering
\includegraphics[width=\columnwidth]{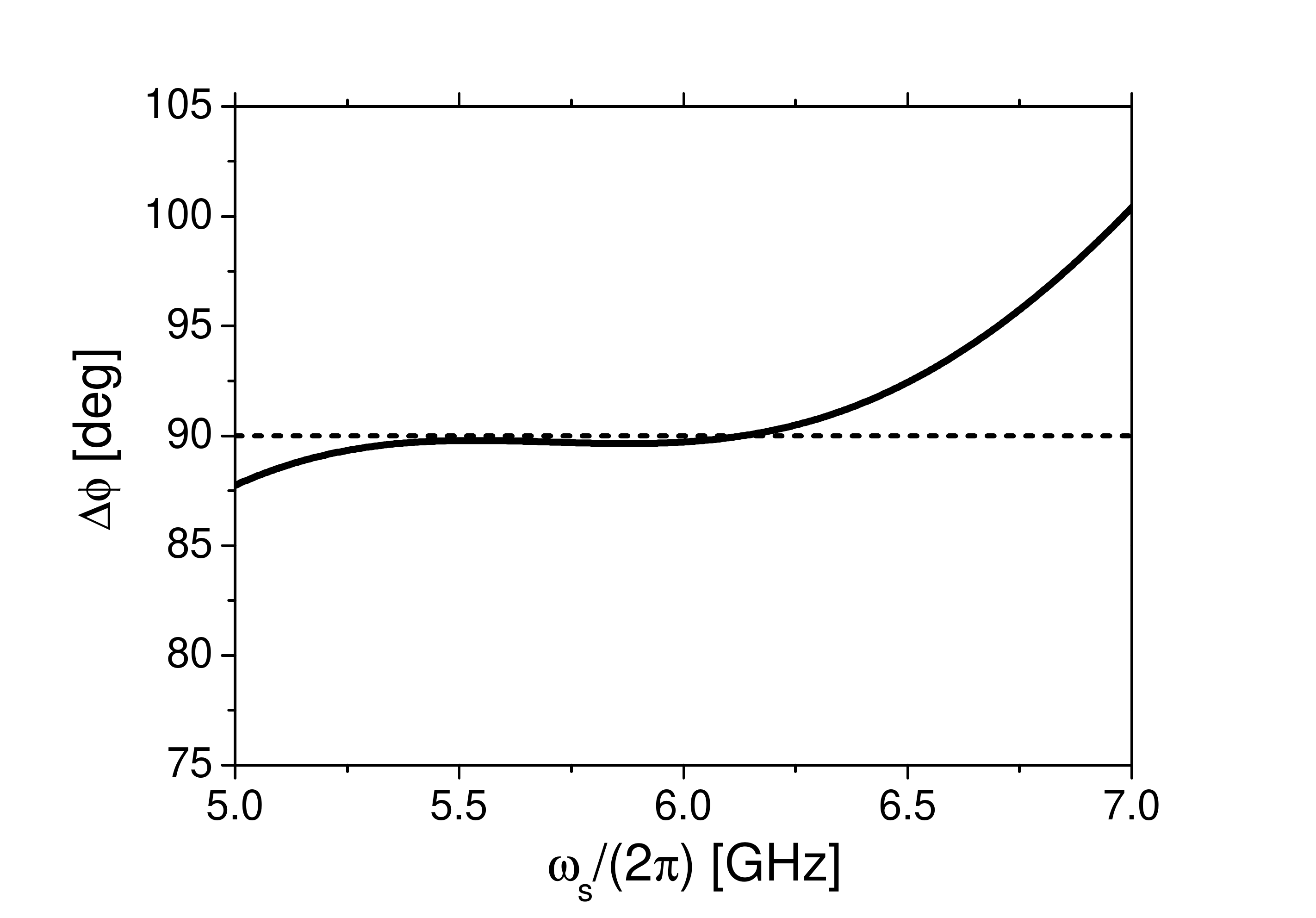}
\caption{\label{figS06}Relative phase difference between the ports (2) and (3) of the 
branch-line coupler in the JPA circuit.}
\end{figure}

\begin{figure}[htb]
\centering
\includegraphics[width=\columnwidth]{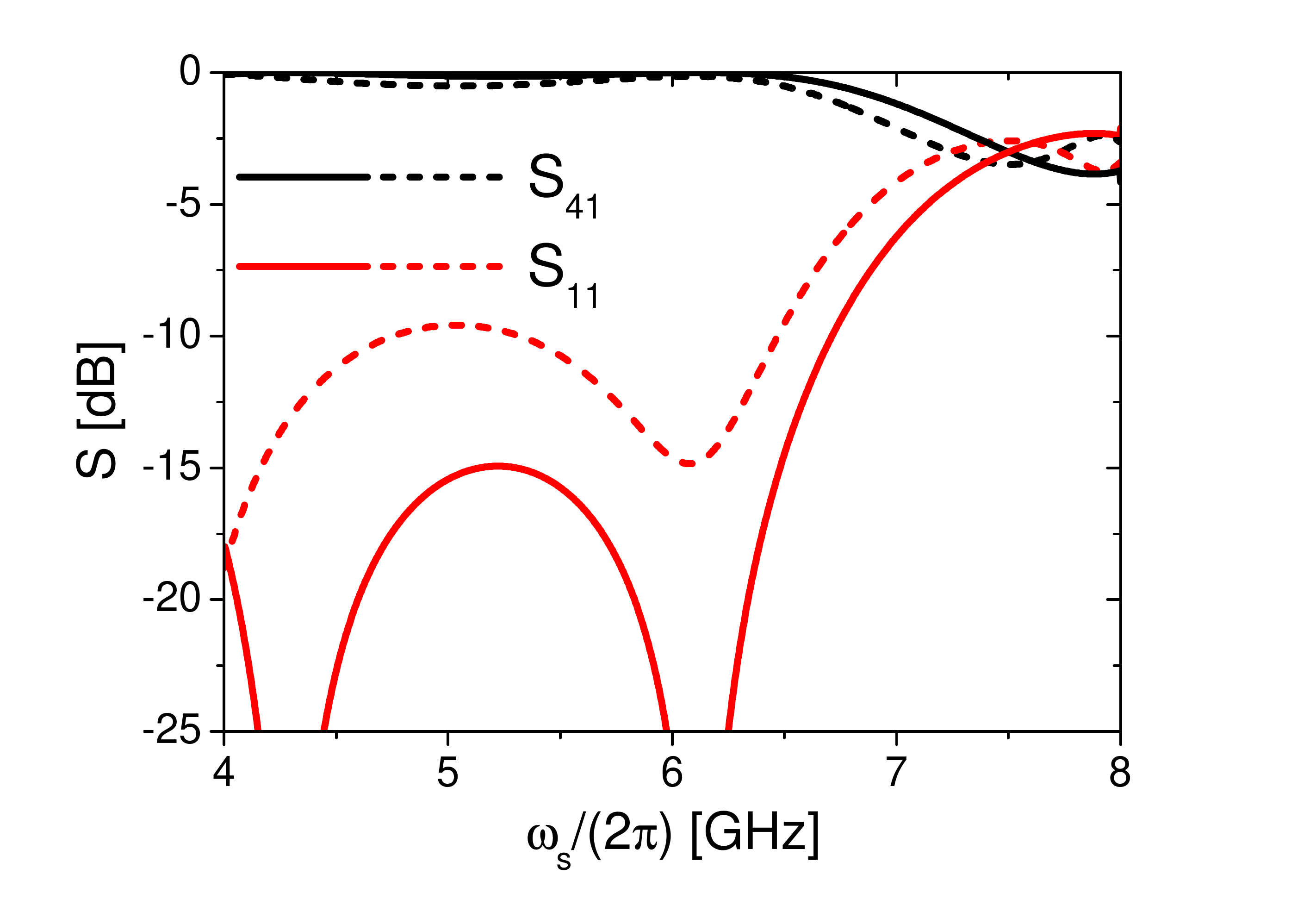}
\caption{\label{figS07}Degraded directivity $S_{41}$ and return loss $S_{11}$ 
as a result of relatively detuned SQUIDs, 
resulting in different plasma frequencies $\omega_{0}$. The incident signals which are 
amplified and reflected from the individual nonlinear JJOs shown 
in Fig.~\ref{fig01}(a) will obtain different phases due to this detuning. 
The figure summarizes the effect of a rather large additional phase difference 
of $\pi/18$ between the two nonlinear JJOs shown as dashed lines, 
compared to the ideal case with equal tuned SQUIDs shown as solid lines.}
\end{figure}

In this section we provide additional circuit characteristics. Figure~\ref{figS05} 
shows the signal coupling from the embedding circuit to the JJOs 
being very close to '1' over the entire operation bandwidth of the 
JPA around 6~GHz. Figure~\ref{figS06} shows the 
relative phase difference between the ports (2) and (3) of the branch-line 
coupler. A stable phase relation of $\pi/2$ over the operation bandwidth of the 
JPA is the basis of the directional signal routing. 
Finally, Fig.~\ref{figS07} shows the effect of relatively detuned SQUIDs 
on the signal routing and return loss, adding 
a phase difference to the amplified 
signals additional to the one imposed by the branch-line coupler.   
\end{document}